\documentclass[10pt]{article}
\pdfoutput=1
\usepackage[mathscr]{eucal}
\usepackage{enumerate}
\usepackage[all]{xy} 
\usepackage{graphicx}
\usepackage{caption,subcaption}
\usepackage[centertags]{amsmath}
\usepackage{amsfonts,amssymb,amsthm}
\usepackage{newlfont,verbatim}
\usepackage{algorithmic,algorithm}
\usepackage[hidelinks]{hyperref}
\usepackage{mathrsfs, mathtools}
\usepackage{bbm} 
\usepackage{geometry}
\usepackage{float}
																									
\theoremstyle{plain}
\newtheorem{thm}{Theorem}[section]

\newtheorem{definition}[thm]{Definition}

\newcommand{\R}{\mathbb{R}}
\newcommand{\E}[1]{\mathbb E\left[#1\right]}
\newcommand{\cov}{\mathrm{Cov}}
\newcommand{\var}{\mathrm{Var}}

\newcommand{\prob}{\mathbb{P}}

\newcommand{\ed}{\overset{d}{=}}
\newcommand{\four}{\mathscr{F}}
\newcommand{\eig}{\mathrm{eig}}
\newcommand{\cim}{\mathrm{cim}}
\newcommand{\riesz}{\mathrm{riesz}}
\title{Power-Law Noises over General Spatial Domains and on Non-Standard Meshes}
\author{Hans-Werner van Wyk, Max Gunzburger, John Burkardt, and Miroslav Stoyanov}

\begin{document}
\maketitle

\begin{abstract}

Power-law noises abound in nature and have been observed extensively in both time series and spatially varying environmental parameters. Although, recent years have seen the extension of traditional stochastic partial differential equations to include systems driven by fractional Brownian motion, spatially distributed scale-invariance has received comparatively little attention, especially for parameters defined over non-standard spatial domains. This paper discusses the generalization of power-law noises to general spatial domains by outlining their theoretical underpinnings as well as addressing their numerical simulation on arbitrary meshes. Three computational algorithms are presented for efficiently generating their sample paths, accompanied by numerous numerical illustrations. 

\end{abstract}

\section{Introduction}

While the results of numerical simulations of physical systems may depend profoundly on the underlying model parameters, these can often only be observed partially and indirectly and arise within complex environments that cannot be described in full deterministic detail. In these cases, parameters are more appropriately modeled as random variables, -processes, or -fields. To be realistic, statistical models of these parameters should ideally not only be consistent with available measurements, but also incorporate broader, more qualitative information. 
Scale-invariance is one such property that has been observed widely both in natural time series, as well as spatially varying random fields. Traditionally, however, random perturbations appearing in stochastic differential equations (SDE's) or stochastic partial differential equations (SPDE's) were almost exclusively modeled as white noise. Composed in equal parts of random fluctuations at all length scales, white noise is relatively simple to generate and well understood, leading to its disproportionate use in light of available experimental evidence. Over the past two decades, a concerted effort was made to extend both the analysis and simulation of solutions of SDE's and SPDE's to include more realistic statistical models for \emph{time-varying} parameters that exhibit scale invariance, such as fractional Brownian motion (fBm) \cite{Biagini2008}. Scale-invariant random parameters over \emph{spatial domains} on the other hand have received comparatively little attention. They are difficult to analyze and simulate, partly due to fact that their covariance function can often not be expressed explicitly. This paper describes the theoretical underpinnings of scale-invariant-, or power-law noises and their numerical approximations, over arbitrary spatial domains, and proposes three computational algorithms to simulate their sample paths over arbitrary meshes. 

\vspace{1em}

Power-law noise refers to a class of statistical signals whose power spectral density $S$, a measure of the power carried by the signal per unit frequency, satisfies the power law $S(\xi)\propto 1/\xi^\alpha$ for some $\alpha\geq 0$ and frequency $\xi>0$ within some range $[\xi_{\min},\xi_{\max}]$. Since the first description of so called `pink noise', following an experiment designed to test Schottky's theory of `shot noise' in vacuum tubes (see \cite{Johnson1925}, \cite{Schottky1926}), power spectral densities exhibiting a power law decay have been observed experimentally in time series related to the voltages of diodes and transistors, the resistance of semiconductors and thin films, the average seasonal temperature, annual rainfall, traffic flows, action potentials in neurons, and the timbre of musical notes, to name but a few (see review articles \cite{Kasdin1995},\cite{Keshner1982},\cite{Milotti2002}). The ubiquitous presence of power-law noise processes in such a wide range of natural- and man-made phenomena has led some researchers to suggest the existence of an underlying, characterizing statistical model for this type of noise that is independent of any specific physical mechanism producing the noisy signal. The development of such a statistical model would not only provide a foundation for understanding power-law noises, but would also contribute to the establishment of numerical algorithms for simulating their sample paths, which in turn can be used in aid of stochastic simulations. In light of the prevalence of power-law noises in nature, replacing more traditional white noise fields with power-law fields is likely to improve the validity of stochastic models and hence their predictive ability \cite{Stoyanov2011}.

\vspace{1em} 

Although a canonical model remains elusive, the past century has seen a wealth of research devoted to the statistical description of power-law noise processes with a variety of different formulations, ranging from Poisson process models \cite{Lowen2005}, fractional Brownian motion \cite{Barnes1966,Mandelbrot1968}, stochastic differential equation models \cite{Kaulakys2006}, fractional differencing and ARFIMA (autoregressive fractional integration moving average) models \cite{Hosking1981}, to approaches based on wavelets \cite{Wornell1993,Flandrin1992} and system theory \cite{Kasdin1995}. A central theme in the literature on $1/\xi^\alpha$-noise is that of scale invariance or statistical self-similarity, referring to the fact that essential statistical features remain unchanged as length scales vary. Indeed, for $1/\xi^\alpha$ power spectral densities, a change in the power per unit frequency due to a scaling of the frequency (i.e. `zooming in') can be undone by simply scaling the signal itself by an appropriate constant. Since scale invariance occurs not only in \emph{time} varying stochastic processes, but can also be observed in numerous \emph{spatially} distributed phenomena, such as landscapes \cite{Milne1992}, glacial surface characteristics \cite{Arnold2003}, fracture formation \cite{Hirata1987} and -surfaces \cite{Bouchaud1997}, interface growth and roughening phenomena \cite{Barabasi1995} (see also Barkhausen noise \cite{Spasojevifmmodecuteclseci1996}), as well as various geophysical structures \cite{Nieto-Samaniego2005}, including subsurface flow and transport parameters \cite{Turcotte1997}, it is useful to extend statistical models of power-law noise to more general domains.    

\vspace{1em}

Strictly speaking, no physical signal obeys the power-law $1/\xi^\alpha$ over its entire frequency spectrum $[0,\infty)$, since this would have unphysical implications for its energy, computed as the integral of its power spectral density over its frequency spectrum. For  values of $\alpha \geq 1$, the energy in a $1/\xi^\alpha$-noise signal defined over the frequency range $[\xi_{\min},\xi_{\max}]$ diverges as $\xi_{\min}\rightarrow 0^+$, as does that in the power-law noise signal with $0\leq \alpha < 1$ when $\xi_{\max}\rightarrow \infty$. In practice, restrictions on both the length of the signal and the sampling frequency, limit physical observations to a frequency interval bounded away from both zero and infinity. Periodogram estimates of $S$ often look flat in the low frequency range, while taking the form $S(\xi)\propto 1/\xi^2$ for high frequencies (see \cite{Milotti2002}), although evidence of the persistence of power laws over large time periods, also known as infra-red divergence, has been reported in \cite{Keshner1982}. In light of the variety of physical systems that exhibit scale invariance, it is therefore important to choose a colored noise model that is both compatible with observations of the system as well as the scientific needs of the modeler. In the simulations of long memory processes such as network traffic trace processes \cite{Willinger1998} for example, it is important for the signal to adhere to the power law $1/\xi^\alpha$ in the low frequency spectrum, while numerical studies of local statistical self-similarity tend to focus on the high frequency range. 
 
\vspace{1em}

\begin{figure}[ht]
\centering
	\begin{subfigure}[t]{0.3\textwidth}
	\includegraphics[width = \textwidth]{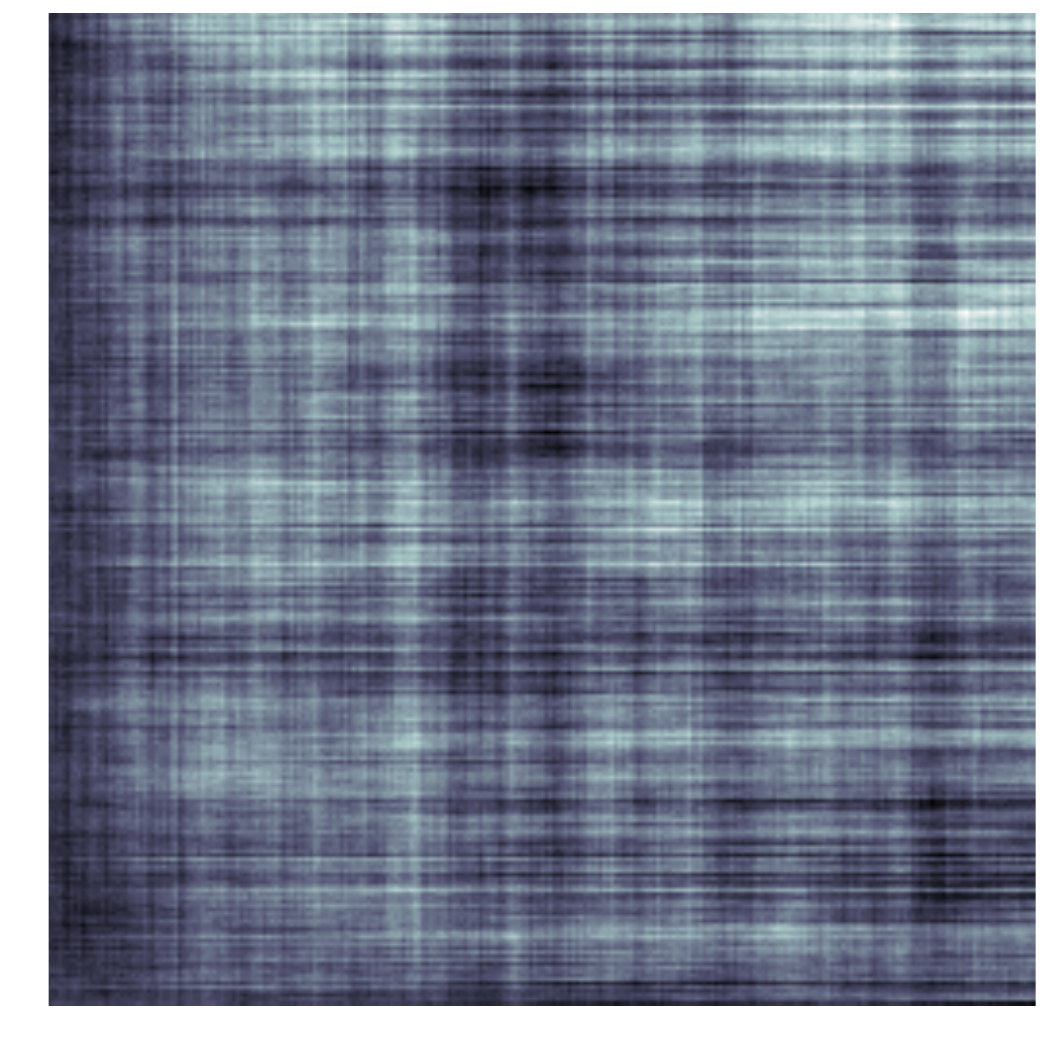}
	\caption{Fractional Brownian sheet over a square with power law $S(\xi_1,\xi_2)= \xi_1^{-\alpha_1}\xi_2^{-\alpha_2}$.}
	\label{subfig:fbm2d_sheet_box}
	\end{subfigure}%
	~	
	\begin{subfigure}[t]{0.3\textwidth}
	\includegraphics[width = \textwidth]{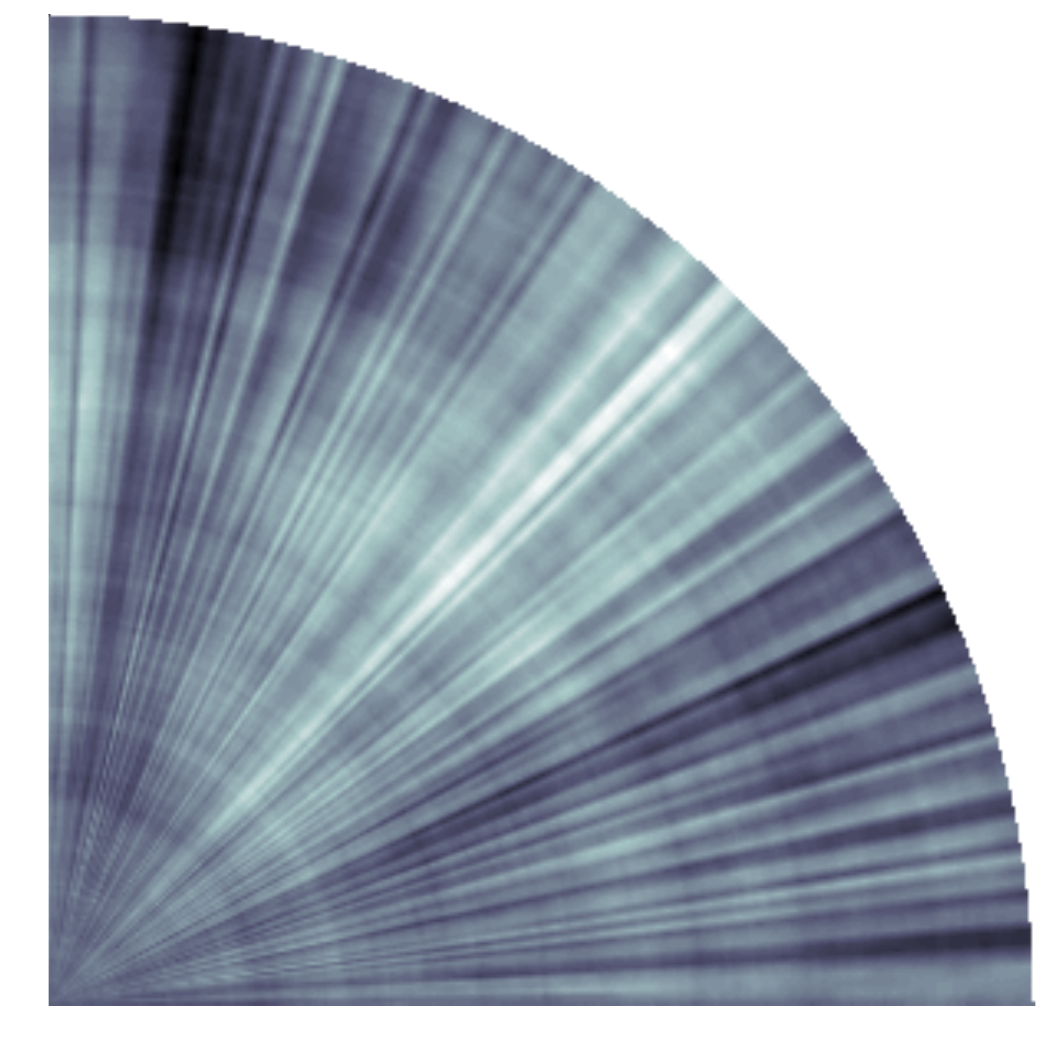}
	\caption{Fractional Brownian sheet over a quarter disc with power law $S(\xi_r,\xi_\theta) = \xi_{r}^{-\alpha_1}\xi_\theta^{-\alpha_2}$.}
	\label{subfig:fbm2d_sheet_polar}
	\end{subfigure}%
	~	
	\begin{subfigure}[t]{0.3\textwidth}
	\includegraphics[width = \textwidth]{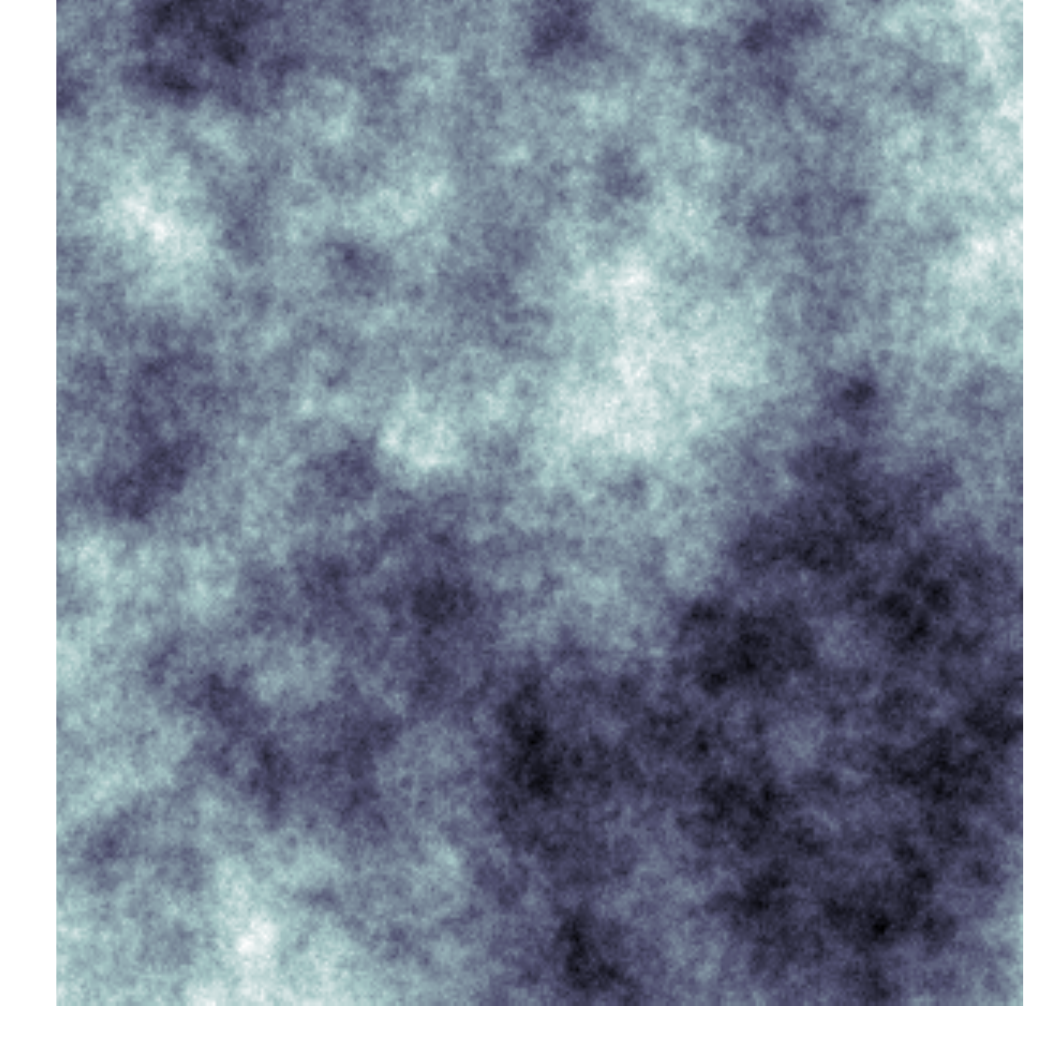}
	\caption{Fractional Brownian surface over a square with power law $S(\xi_1,\xi_2)=\|\xi\|^{-\alpha}$.}
	\label{subfig:fbm2d_surface_box}
	\end{subfigure}
	\caption{Different power law noises over two dimensional domains}			  
	\label{fig:2d_power_laws}
\end{figure}

This paper discusses the numerical simulation of power-law noise fields over general multidimensional regions. We restrict ourselves to centered Gaussian noise fields, since these are used widely in practice and are uniquely determined by their covariance function. There are multiple possibilities for defining power laws over higher dimensional frequency domains, some of which are illustrated in Figure \ref{fig:2d_power_laws}. The field may obey a different power law in each of its components (see Figures \ref{subfig:fbm2d_sheet_box} and \ref{subfig:fbm2d_sheet_polar}), or the power law may be written in terms of the frequency vector's radial distance from the origin (see Figure \ref{subfig:fbm2d_surface_box}). Other forms are also possible. The fundamental challenge in generating power-law noises, even in one dimension, stems from the difficulty of translating requirements on the signal's power spectral density, i.e. the requirement $S(\xi)\propto 1/\xi^\alpha$, into quantifiable properties that are useful for construction of the signal, in this case the form of the covariance. This is partly due to the fact that different values of $\alpha$ give rise to signals with widely divergent statistical properties. While the Wiener-Khinchine Theorem \cite{Wiener1930, Khintchine1934} relates auto-covariance functions of stationary processes to their power spectral densities via Fourier transform, this assumption does not hold for all $\alpha\geq 0$. For an intuition of the correlation structure of a general power-law noise process, it is useful to consider again the signal's energy. For small values of $\alpha$, high frequencies contribute significantly to the total energy, suggesting erratic, uncorrelated behavior and hence stationarity, the extreme case of which is white noise ($\alpha = 0$) where all frequencies contribute equally to the total energy. As $\alpha$ increases, the contribution of the high frequency components is diminished, compared to that of the low frequencies and the signal tends to become smoother, more correlated, as well as non-stationary (see Brownian noise, $\alpha = 2$).

\vspace{1em} 

\begin{figure}[ht]
	\centering
    \begin{subfigure}[t]{0.3\textwidth}
    		\includegraphics[width=\textwidth]{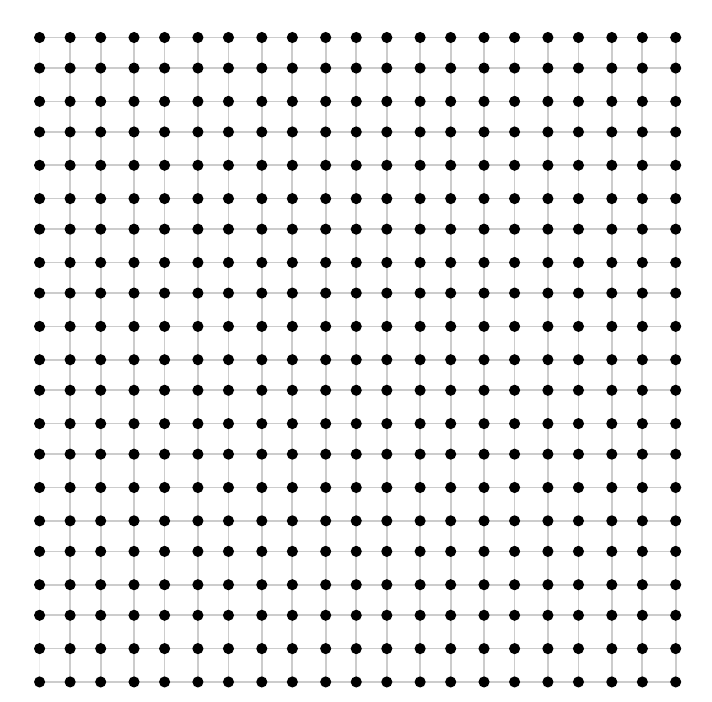}
        \caption{A regular grid on a rectangular region}
        \label{subfig:square_grid}
    \end{subfigure}%
    ~
    \begin{subfigure}[t]{0.3\textwidth}
    		\includegraphics[width=\textwidth]{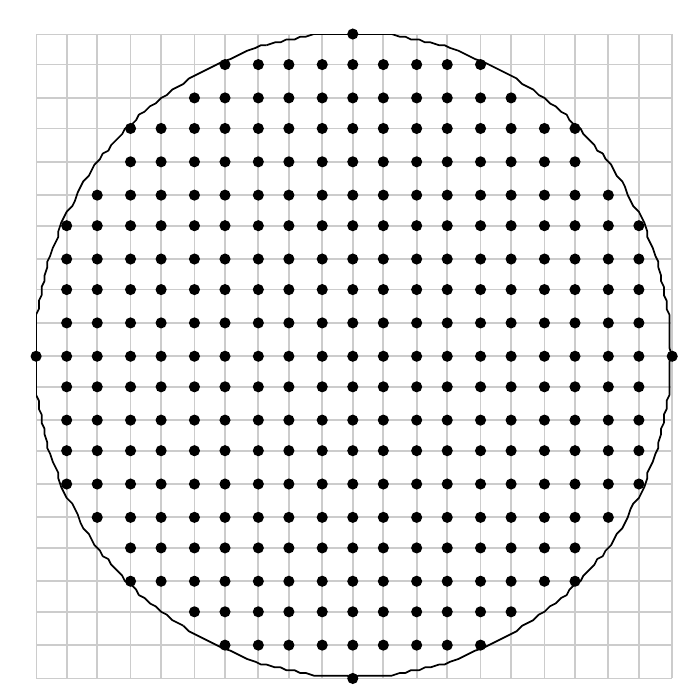}
    		\caption{A regular grid on a circular region}
    		\label{subfig:circle_grid}
    		\end{subfigure}%
    ~ 
    \begin{subfigure}[t]{0.3\textwidth}
    		\includegraphics[width=\textwidth]{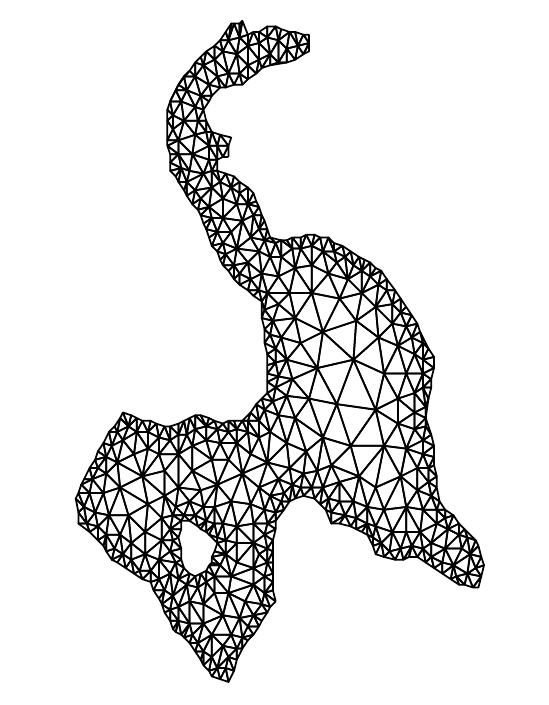}
        \caption{A finite element mesh over a general region}
        \label{subfig:lake_grid}
    \end{subfigure}%
    \caption{Colored noise can be synthesized efficiently on regular grids over simple domains. However, non-uniform meshes over general domains are often preferred.}\label{fig:grids_domains}
\end{figure}

Considering the above-mentioned subtleties, many colored noise models are formulated by directly specifying the covariance structure of the field, instead of deducing its form based on specifications of the power law. Fractional Brownian motions, as well their multi-dimensional analogues, fractional Brownian surfaces and -sheets, form a widely used class of such noise models. In a sense they can be said exhibit power law behavior \cite{Flandrin1989}, in addition to having other desirable properties, such as H\"older continuous sample paths and stationary increments. Moreover, efficient algorithms, such as the circulant embedding method \cite{Chan1998,Dietrich1997}, have been adapted \cite{Stein2002} to generate fast, exact numerical simulations of fractional Brownian surfaces on equispaced rectangular grids defined over certain standard domains, such as rectangles or circles (see Figures \ref{subfig:square_grid} and \ref{subfig:circle_grid}). The covariance function of the standard fractional Brownian surface, however, does not depend on the region's underlying geometry, which can lead to unrealistic models for fields defined over non-convex regions (see Figure \ref{subfig:lake_grid} and Subsection \ref{subsection:fbm}). Moreover, if the field represents the input for a complex physical system that needs to be solved numerically, the meshes imposed on the region are often non-uniform for the sake of computational expediency. Depending on the application, it may also be more appropriate for the field to be stationary, in which case the power law no longer holds in the low frequency range. Elliptic Gaussian fields \cite{Benassi1997} generalize fractional Brownian motion, based on its spectral characterization in terms of the fractional Laplace operator. They are self-similar random fields (at least locally) and  have H\"older continuous sample paths. Furthermore, this theoretical framework allows for the definition of colored noise fields with prescribed values in certain regions, and has even been used to define colored noise fields over manifolds (\cite{Gelbaum2012, Gelbaum2013}), by means of the Laplace-Beltrami operator. 

\vspace{1em}

After some preliminary remarks and establishing notation in Section \ref{section:preliminaries}), we review fractional Brownian surfaces and their generalizations, the elliptic Gaussian fields over Euclidean space $\R^d$ in Section \ref{section:elliptic_fields}. Section \ref{section:riesz_fields} treats the Riesz fields, generalizations of elliptic Gaussian fields to arbitrary bounded regions. The numerical simulation of Riesz fields is the subject of Section \ref{section:samplepaths}. We discuss numerical simulations based on the approximation of the discretized fractional Laplace operator, either through the eigen-decomposition of the finite element Laplace matrix or the contour integral method \cite{Burrage2012}, as well as simulations based on a modified form of the Riesz potential, which allows for the simulation of multi-fractional fields for which the level of roughness may vary throughout the domain. Numerical illustrations accompany the discussion throughout. For the sake of visualization, all of our computational results are based on regions in $\R^2$. The algorithms discussed, however, extend readily to arbitrary dimensions. Finally, Section \ref{section:conclusion} contains concluding remarks. Considering the vastness of this research field, it is inevitable that this paper omits many important approaches, such as those based on wavelet approximations. In \cite{Wornell1993} (see also \cite{Flandrin1992}), the author cogently argues for the appropriateness of wavelets as a framework for analyzing power-law noises.

\section{Notation and Preliminaries}\label{section:preliminaries}

Let $(\Omega,\mathcal F, \prob)$ be a complete probability space and $D\subset \R^d$ a bounded region acting as the index set for our random field. Throughout, we use $X$ and $Y$ to denote generic random fields, while $\mathbf{X}$ and $\mathbf{Y}$ represent random vectors. A Gaussian random field $\{X(x)\}_{x\in D}$ is a collection of random variables so that for any finite subset $\{x_i\}_{i=1}^n\subset D$, the vector $\mathbf X = [X(x_1),...,X(x_n)]^T$ has a joint Gaussian distribution. A random field is said to be centered if its expectation $\E{X(x)}=0$ for all $x\in D$, an assumption we make throughout for the sake of simplicity. The covariance function $C_X:D\times D \rightarrow \R$ of a centered random field $X$ is defined pointwise for any two points $x$ and $y$ by the covariance $C_X(x,y):=\cov(X(x),X(y))=E[X(x)X(y)]$. Fractional Brownian motions as well as -surfaces with Hurst parameter $H$ are denoted by $B_H$.  White noise fields $B_0$ are mostly denoted by $W$, especially when referring to the isotropic white noise process. The random vector $\mathbf{Z}=[Z_1,...,Z_n]^T$ always represents a vector of identically distributed (i.i.d.) Gaussian random variables.  To prevent unnecessary ambiguity, we use the variables $t$ or $s$ to index stochastic processes and $x$ or $y$ to index random fields over higher dimensional regions.  

\vspace{1em}

The power spectral density of a deterministic signal $\{x(t)\}_{t\geq 0}$ at a given frequency $\xi$, is simply the magnitude of its Fourier transform squared, i.e. $S(\xi)=|\four(x)(\xi)|^2$, where $\four(x)(\xi):=\int_0^\infty x(t)e^{-i t\xi}\;dt$. Since the Fourier transform of a random signal $\{X(t)\}_{t\geq 0}$ over $[0,\infty)$ does not exist for many signals of practical importance, the power spectral density is often defined via the truncated Fourier transform   
\[
\four_T(X)(\xi,\omega):=\frac{1}{\sqrt{T}}\int_0^T X(t,\omega)e^{-i\xi t}\;dt,\ \ \text{for } \omega \in \Omega \ \text{ and } \xi \geq 0,
\]  
and hence
\[
S(\xi) := \lim_{T\rightarrow \infty} \E{|\four_T(X)(\xi)|^2}.
\]
In our numerical calculations, we approximate the power spectrum by simply computing the discrete Fourier transform of its sample paths over a finite domain, using fast Fourier transforms, and averaging the square of their magnitudes. In multi-dimensional domains in $\R^d$, we use the multi-dimensional fast Fourier transform.   

\vspace{1em}

For a stationary time series $\{X(t)\}_{t\in\R}$, the Wiener-Khintchine Theorem can be used to directly relate its power spectrum to its auto-covariance function $\rho(\tau):=\cov(X(t),X(t+\tau))$ via the Fourier transform, i.e. $S(\xi) = \hat{\rho}(\xi):=\int_{-\infty}^\infty \rho(\tau)e^{-i\xi \tau}\;d\tau$. Although no equivalent relation exists for non-stationary processes $\{X(t)\}_{t\in \R}$, the time dependent Wigner-Ville spectrum
\[
S^{WV}_X(t,\xi) := \int_{-\infty}^{\infty} \cov\left(X\left(t+\frac{\tau}{2}\right),X\left(t-\frac{\tau}{2}\right)\right)e^{-i\xi \tau}\; d\tau,
\]
presents a convenient generalization. In \cite{Flandrin1989}, this version of the power spectrum is used to show that fractional Brownian motion is a $1/\xi^\alpha$-process. 

\vspace{1em}

It is often convenient to define a Gaussian random field as the convolution of a deterministic kernel function with white noise. To make this construction rigorous, we make use of the Skorokhod integral with respect to the isotropic white noise process $W$, defined on Hilbert space $H$ as the mapping $W:H \rightarrow L^2(\Omega)$ so that i) for any $h\in H$, $W(h)\sim N(0,\|h\|^2)$ and ii) for any $h_1,h_2\in H$, $\E{W(h_1)W(h_2)} = \langle h_1,h_2\rangle_H$ (see \cite{Nualart2006}). Let $\mathcal S$ be the set of $H$-valued random variables of the form $F = f(W(h_1),...,W(h_n))$, where $h_1,...,h_n\in H$ and $f:\R^n \rightarrow \R$ is infinitely differentiable with partial derivatives that grow at most polynomially at infinity. Further, denote by $D(F)$ the Malliavin derivative of $F$, i.e. $D(F)= \sum_{i=1}^n \partial_i f(W(h_1),...,W(h_n))h_i$, which can be extended to a closed unbounded operator $D:L^2(\Omega)\rightarrow L^2(\Omega,H)$ with domain $\mathbb D^{1,2}$, the Sobolev-Watanabe space. Since $\mathbb{D}^{1,2}$ is dense in $L^2(\Omega)$, there exists a unique adjoint operator $\delta:\mathrm{dom}(\delta)\rightarrow L^2(\Omega)$ defined for each $u\in \mathrm{dom}(\delta)$ by the relation
\[
\E{F\delta(u)} = \E{\langle D(F),u\rangle_H}\ \ \text{for all } F\in \mathbb D^{1,2},
\]
where $\mathrm{dom}(\delta) = \{u\in L^2(\Omega,H): F\mapsto \E{\langle D(F),u\rangle_H} \text{ is bounded }\}$. This operator is also known as the Skorokhod integral, written as $\int u \;dW := \delta(u)$. For deterministic integrands $u\in H$, the isometry $\|\int u \;dW\|_{L^2(\Omega,H)} = \|u\|_H$ allows these integrals to be approximated by integrals  of simpler functions. Let $V^n\subset H$ be a finite dimensional subspace, spanned by basis functions $\{\phi_i\}_{i=1}^n$. Then 
\[
\int \sum_{i=1}^n c_i \phi_i\;dW = \sum_{i=1}^n c_i Z_i, \ \text{where } [Z_1,...,Z_n] \sim N(0,\Sigma), \ \text{with } \Sigma_{ij} = \langle \phi_i,\phi_j\rangle_H
\]
for $i,j=1,...,n$. In particular, when this basis consists of simple functions over a bounded domain, then the isotropic white noise process $W$ takes the form of the white noise measure, defined on the set of Borel measurable subsets of $\R^d$ with finite volume (see \cite{Janson1997}). For any such Borel set $A$ with volume $|A| <\infty$, $W(A)\sim N(0,|A|)$, and for any finite collection $A_1,...,A_n$ of disjoint sets, the $W(A_i)$ are independent and $W(\cup_{i=1}^n A_i) = \sum_{i=1}^n W(A_i)$. In this case, the white noise integral of a simple function $f_s(x):=\sum_{i=1}^{n}c_i \chi_{A_i}(x)$ $\int f_s(x)\;dW(x) := \sum_{i=1}^n c_i W(A_i)$.
 

\section{Models for Power-Law Noises over $\R^d$}\label{section:elliptic_fields}

In certain simple cases, statistical models for power-law noises can be developed directly, based on specifications of their desired properties. In this section, we introduce the fractional Brownian motion (fBm) over the interval $[0,\infty)$ as the unique self-similar Gaussian process with stationary increments. Although the self-similarity required by Definition \ref{definition:fBm} does not involve the power spectral density, it can be shown that the Wigner-Ville spectra of fBm's do indeed exhibit a power law decay. The covariance function of fBm can directly be generalized to Euclidean space, giving rise to the fractional Brownian surface (fBs) over $\R^d$. This extension is too rigid, however, rendering fBs unsuitable to model spatially correlated noise in many practical applications. Their spectral characterization nevertheless identifies them as members of the family of elliptic Gaussian fields, Gaussian fields defined in terms of pseudo-differential operators with positive symbol. Using these operators (and in particular the fractional Laplace operator) as the basis for the generalization of power-law noises gives rise to models suitable for arbitrary domains that share the salient properties of fBs, such as H\"older continuity, while also allowing for additional flexibility, such the imposition of boundary conditions.

\subsection{Fractional Brownian Motion}\label{subsection:fbm}

Fractional Brownian motion, a family of Gaussian random processes parameterized by the Hurst parameter $H\in (0,1)$, has become a particularly widespread model for time dependent power-law noises. In the seminal paper \cite{Mandelbrot1968}, the authors introduce scale invariance directly through the definition of self-similarity with respect to a Hurst parameter $H$. 

\begin{definition}\label{definition:self_similarity_global}
A random process $X:[0,\infty) \rightarrow \R$ is said to be self-similar with Hurst parameter $H \in (0,1)$ (H-s.s.) if for every $c>0$, we have 
\[
X(ct) \ed c^{H}X(t), \ \ \text{where `$\ed$' denotes equality in distribution}.
\] 
\end{definition}

This definition generalizes the well-known self-similarity property of Brownian motion, whose Hurst parameter $H=1/2$. For any $t\in [0,\infty)$, let $\triangle X(t,h) = X(t+h) - X(t)$ denote the $h$-increment process related to $X$. If in addition to H-s.s., we assume that $X(0)=0$ almost surely (a.s.) and that the increments $\triangle X(t,h)$ are stationary (i.e. the distribution of the increment depends only on $h$), then we automatically arrive at the following explicit form for its covariance function, namely 
\begin{align*}
\E{X(t)X(s)} &= \frac{1}{2} \left(\E{X(t)^2} + \E{X(s)^2} - \E{\triangle X(s,t-s)^2}\right)\\
&= \frac{1}{2} \left(\E{X(t)^2} + \E{X(s)^2} - \E{(X(t-s))^2}\right)\\
&= \frac{1}{2} \left(t^{2H}\E{X(1)^2} + s^{2H}\E{X(1)^2} - (t-s)^{2H}\E{X(1)^2}\right)\\
&= \frac{\E{X(1)^2}}{2} (t^{2H} + s^{2H} - |t-s|^{2H}),
\end{align*}
for any points $s,t\in [0,\infty)$. The above equations also serve to prove the converse, namely that any zero mean stochastic process with this covariance matrix necessarily has stationary increments and is H-s.s.. For simplicity, we assume henceforth that $\E{X(1)^2} = 1$. The definition of fractional Brownian motion is given by the following.%
\begin{definition}[Fractional Brownian Motion]\label{definition:fBm}
A fractional Brownian motion $B_H(t)$ with Hurst parameter $H\in(0,1)$ is a continuous and centered Gaussian process, i.e. $B_H(0)=0$ and $\E{B_H(t)}=0$ for $t\geq 0$, with covariance function given by
\begin{equation}
C_{B_H}(s,t) := \E{B_H(s)B_H(t)} = \frac{1}{2}\left(s^{2H} + t^{2H}-|t-s|^{2H}\right), \ \ t,s \geq 0.\label{eqn:fbm_covariance}
\end{equation}%
\end{definition}%
\noindent Although fractional Brownian motion, unlike standard Brownian motion, is neither mean square differentiable nor a martingale for $H\neq \frac{1}{2}$, it does admit a H\"older continuous modification \cite{Biagini2008}. Moreover, its variance $\var(B_H(t))= \E{B_H(t)^2}=t^{2H}$ for $t\geq 0$, fBm is non-stationary. 

\vspace{1em}

\begin{figure}[h]
\centering
\includegraphics[width = 0.5\textwidth]{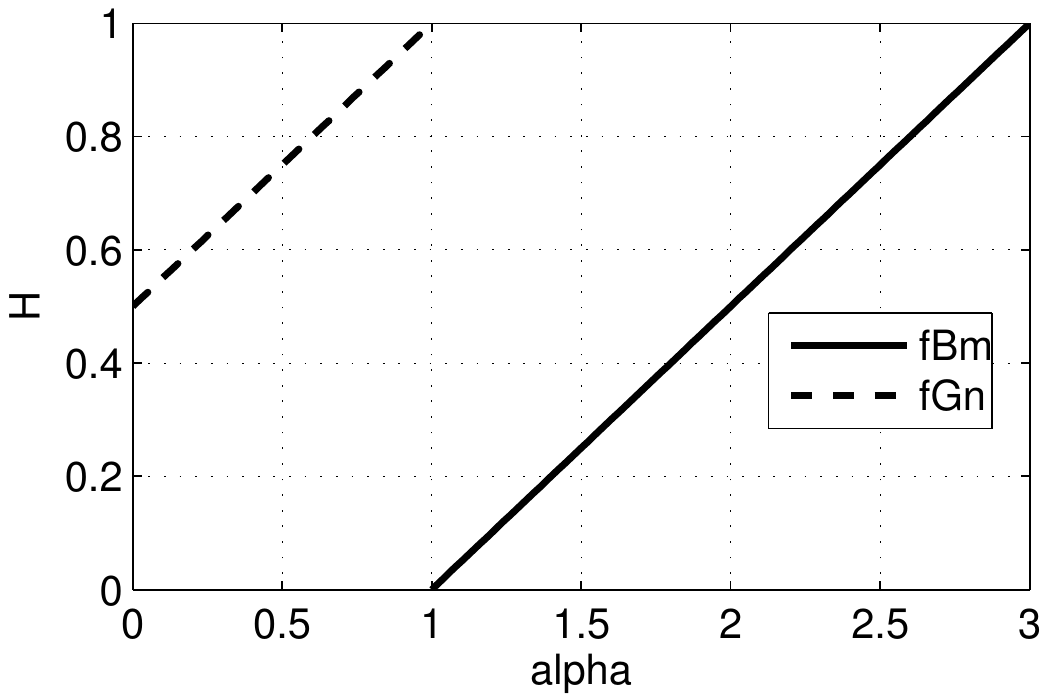}
\caption{Relation between the power $\alpha$ and Hurst parameter $H$ for fractional Brownian motion and fractional Gaussian noise. Note that no value of $H\in (0,1)$ corresponds to $1/f$-noise.}\label{fig:alpha_vs_Hurst}
\end{figure}

In \cite{Flandrin1989}, it was shown that the power spectrum $S_{B_H}(\xi)$ of fractional Brownian motion, defined in terms of the Wigner-Ville spectrum, satisfies $S_{B_H}(\xi)\propto 1/\xi^{2H+1}$, suggesting that the Hurst parameter $H$ can be related to the power $\alpha >0$ via $\alpha = 2H + 1$. A Hurst parameter $H\in (0,1)$ thus gives rise to power-law noises with $\alpha \in (1,3)$. Colored noise with $\alpha \in [0,1)$ can also be defined by making use of the increment process $\triangle B_H(t,h)$, also known as fractional Gaussian noise (fGn). In fact, it can be shown that the power spectrum of the increment process satisfies $S_{\triangle B_H}(\xi) \propto 1/\xi^{2H-1}$, independent of $t$, so that fractional Gaussian noises are power-law noises with $\alpha\in [0,1)$. The relation between the Hurst parameter $H$ and the power $\alpha$ for fBm and fGn are summarized in Figure \ref{fig:alpha_vs_Hurst}.

\vspace{1em}

The fGn process, while still correlated in general, is nevertheless stationary. Indeed, for $t,h\geq 0$ and $s = t + nh\geq 0$ for some $n\geq 0$, it can be shown that 
\begin{align*}
\cov	(\triangle B_H(t,h),\triangle B_H(s,h)) &= \E{\triangle B_H(t,h)\triangle B_H(s,h)}\\
 &= \frac{1}{2}h^{2H}[(n+1)^{2H}+(n-1)^{2H}-2n^{2H}], 
\end{align*}
which is independent of $s,t \geq 0$.

\vspace{1em}

The covariance function defined in \eqref{eqn:fbm_covariance} readily suggests an extension of fractional Brownian motion over the index set $\R^d$.%
\begin{definition}[Fractional Brownian Surface] A fractional Brownian surface $\{B_H(x)\}_{x\in \R^d}$ with Hurst parameter $H\in (0,1)$ is defined to be the continuous, centered Gaussian random field with covariance function
\begin{equation}\label{eqn:fbf_covariance}
C_{B_H}(x,y) = \E{B_H(x),B_H(y)} := \frac{1}{2}\left( \|x\|^{2H} + \|y\|^{2H}-\|x-y\|^{2H}\right)  
\end{equation}%
\end{definition}
\noindent The fractional Brownian surface retains most of the attractive features of fractional Brownian motion. It is self-similar in the sense that the field $\{B_H(c x)\}_{x\in\R^d}$ agrees with $\{c^H B_H(x)\}_{x\in\R^d}$ in law, for any scaling factor $c > 0$. Moreover, the covariance function \eqref{eqn:fbf_covariance} determines a probability measure over the set of H\"older continuous functions of degree $H$.

\vspace{1em}

\begin{figure}[ht]
	\centering
	\begin{subfigure}[t]{0.4\textwidth}
	\includegraphics[width=\textwidth]{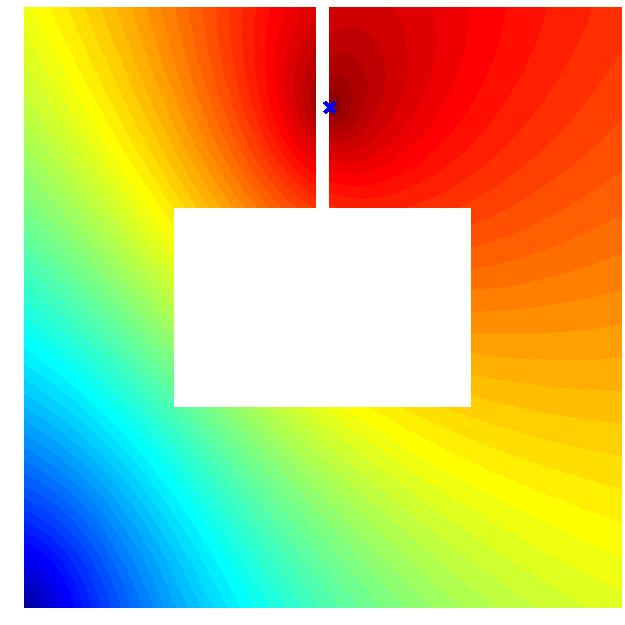}
	\caption{Fractional Brownian Surface}
	\end{subfigure}%
	\hspace{2em}
	\begin{subfigure}[t]{0.4\textwidth}
	\includegraphics[width=\textwidth]{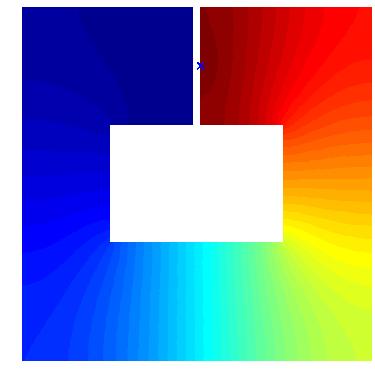}
	\caption{Riesz random field}
	\end{subfigure}
	\caption{The covariance of colored noise fields at a point (cross).}
	\label{figure:covariance_geometry}
\end{figure}

Some properties of this power-law noise model, however, make it unsuitable for certain applications over general regions. Firstly, fractional Brownian fields are non-stationary, like fractional Brownian motion, with a variance that is zero at a point of origin $x_0\in\R^d$ (usually $x_0=0$) and growing at the rate $\|x-x_0\|^{2H}$ as $x\in \R^d$ moves away from $x_0$. This seems natural for stochastic processes, where the origin signifies the present time when the process is known, but it may no longer be relevant in the spatial domain, where there may be no points, multiple points, or even whole regions, in which the field is known exactly (see e.g. \cite{Cressie1985,Krige1951} and Figure \ref{subfig:fbm2d_plus_interior_bnd}). Secondly, while the function given by \eqref{eqn:fbf_covariance} could plausibly be employed to measure the covariance between points in $\R^d$, or even points in a convex sub-domain, it cannot capture the covariance structure of fields defined over general non-convex regions, such as the one depicted in Figure \ref{figure:covariance_geometry}, where there is no longer a direct correspondence between the length of the shortest path between two points and their Euclidean distance.

\vspace{1em} 

Physically accurate colored noise models over general regions should therefore somehow incorporate the geometry of the underlying index set $D$. Similar issues arise in defining  fractional Brownian fields over manifolds. One approach \cite{Istas2005} is to modify the covariance function in \eqref{eqn:fbf_covariance} by replacing the Euclidean norm $\|x-y\|$ with a geodesic distance $d(x,y)$ for $x,y\in D$. It then remains to show that the resulting covariance function is positive definite, by proving for example that the metric $d(x,y)$ is of negative type. Here we follow another approach, based on the spectral characterization of fractional Brownian noise in terms of the Laplace operator on $D$. We begin by discussing the generation of colored noise through the fractional integration of white noise.

\subsection{Hosking's Fractional Difference Model}
 
In \cite{Hosking1981}, the author introduces a discrete version of power-law noise by analogy with the random walk approximation $\{B_n\}_{n=0}^\infty$ of Brownian motion $\{B(t)\}_{t\geq 0}$, given by the convolution sum 
\[
B_n = \sum_{i=0}^n h_i Z_{n-i}, \ n = 1,2, ... ,
\]
where $Z_i\sim N(0,1)$ are identically distributed (iid) standard normal random variables and the discrete impulse response $h_i = 0$ for $i=0$ and $h_i = 1$ for $i=1,2,...,n$. If we let $L$ denote the lag operator, i.e. $L B_n = B_{n-1}$ for $n=1,2,...$ and $L B_0 = B_0 = 0$, then the first order difference satisfies $(1-L)B_n = B_n - B_{n-1} = Z_n$ for $n=1,2,...$. A power-law noise signal $\{B_n^{\beta}\}_{n=0}^\infty$ can now be defined as an ARFIMA (AutoRegressive Fractional Moving Average) process by specifying that its $\beta^{th}$ fractional order difference is a white noise process, i.e. 
\[
(1-L)^{\beta} B_n^{\beta} := \sum_{k=0}^\infty \binom{\beta}{k}(-L)^k B_n^{\beta} = Z_n, \ \text{for } n = 1,2,...\ \text{.}
\]
The process $B_n^{\beta}$ itself can therefore be regarded as a type of fractional cumulative sum of the white noise process $Z_n$.  
To obtain an explicit description of $B_{n}^{\beta}$ as a discrete convolution with a white noise process, we note that the unilateral Z-transform $H^{\beta}$ of its impulse response vector $\{h_i^{\beta}\}_{i=0}^\infty$ must be of the form $H^{\beta}(z) = \frac{1}{(1-z^{-1})^{\beta}}$, with $z>1$. In \cite{Kasdin1995} (see also \cite{Stoyanov2011}), the author  arrives at the same 
transfer function $H^{\beta}$, by `interpolating' between the transfer function of white noise ($\alpha=0$), given by $H(z)=1$, and that of Brownian noise ($\alpha=2$), given by $H(z) = \frac{1}{1-z^{-1}}$. The parameter $\beta$ is thus related to $\alpha$ via $\beta = \alpha/2$. The discrete impulse response function $h_n^{\beta}$ can be computed by means of a simple recurrence relation and sample paths with the right power spectral decay can be generated efficiently through the use of fast Fourier transform. 

\vspace{1em}

\begin{figure}[ht]
	\centering
	\begin{subfigure}[t]{0.45\textwidth}
	\includegraphics[width=\textwidth]{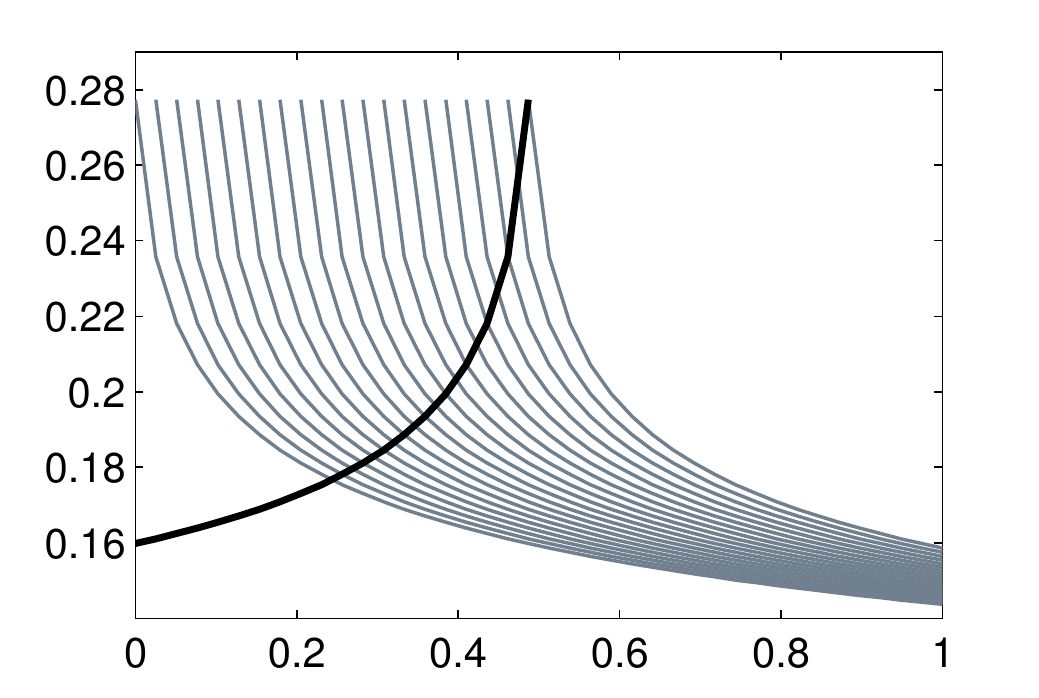}
	\caption{Hosking Process ($\beta=0.85$).}
	\end{subfigure}
	~
	\begin{subfigure}[t]{0.45\textwidth}
	\includegraphics[width=\textwidth]{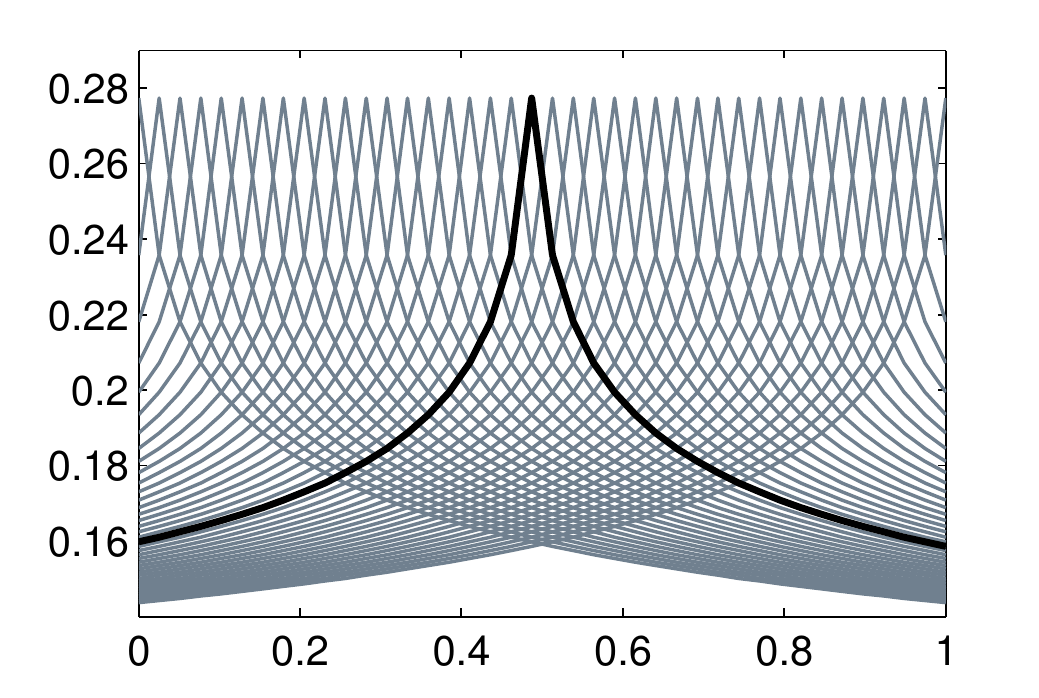}	
	\caption{Modified Hosking Process ($\beta=0.85$)}
	\end{subfigure}
	\caption{Schematic of the composition of a Hosking noise signal. The gray curves represent the influence of each white noise source term over the interval. The black curve attributes the contribution of the various noise sources to the signal's value at $t=0.5$.}\label{fig:hoskings_noise_schematic}
\end{figure}

The value of a Hosking colored noise process at any point of time $t$ is thus determined by a collection of noise sources spread out over the domain (see Figure \ref{fig:hoskings_noise_schematic}), whose influence decreases as their distance to $t$ increases. Lower values of $\alpha$ imply steeper decline in their influence (indicated by the gray curves in \ref{fig:hoskings_noise_schematic}), indicating that the influence of the noise terms are more local, while larger values of $\alpha$ allow the noise sources to have wider ranging influence. In anticipation of simulations of spatially varying noise, Figure \ref{fig:hoskings_noise_schematic} b) shows a modification of the Hosking process in which the noise sources are positioned on both sides of the current time point.  

\subsection{Elliptic Gaussian Fields}

The idea of forming colored noise as the fractional integral of white noise dates back to 1953, when Paul L\'evy \cite{Levy1953} (see also \cite{Barnes1966}) commented on a  stochastic process formed by the cumulative fractional Riemann-Liouville integral of white noise, 
\[
X(t) = \frac{1}{\Gamma(H+\frac{1}{2})} \int_0^t (t-s)^{H-\frac{1}{2}}dW(s),
\]
where $H$ may be any positive number. The Riesz potential $I_s$, defined for any locally integrable function $f:\R^d\rightarrow \R$ by
\begin{equation}\label{eqn:riesz_potential}
I_s[f](x) := c_s \int_{\R^d} \frac{f(y)}{\|x-y\|^{d-s}}\;dy, \ \text{ where }  c_s = \frac{\Gamma(\frac{d-s}{2})}{\pi^{\frac{d}{2}}2^s\Gamma(\frac{s}{2})},\ \ s \in (0,d)
\end{equation}
is the multi-dimensional generalization of the Riemann-Liouville integral. The Fourier transform of the Riesz kernel, together with the convolution theorem reveal that
$\four(I_s f)(\xi) = \|2\pi\xi\|^{-s} \four(f)(\xi)$ and hence $I_s$ is a Fourier multiplier. In comparison, the Laplace operator satisfies 
$\Delta e^{2\pi i x\cdot \xi} = -4\pi^2\|\xi\|^2 e^{2\pi i x\cdot \xi}$,
so that $\four(\Delta f)(\xi) = -4\pi^2\|\xi\|^2 \four(f)(\xi)$. In a spectral sense, the Riesz potential therefore represents the fractional inverse $(-\Delta)^{-\frac{s}{2}}$ of the negative Laplace operator over $\R^d$. %
%
%
It is this connection with the fractional Laplace operator that allows for the development of models for power-law noise fields that preserve the essential properties of Euclidean fractional Brownian fields, such as scale invariance and the stationarity of its increments under translations and rotations, while also reflecting the underlying geometry of the index set.

\vspace{1em}

More generally, let $\mathcal A$ be a pseudo-differential operator defined on an appropriate Hilbert space $\mathcal H_{\mathcal A}$ of functions in terms of its Fourier transform, i.e.
\[
(\mathcal A f)(x):=\frac{1}{(2\pi)^{\frac{d}{2}}}\int_{\R^d} e^{ix\cdot\xi}\sigma(x,\xi)\four(f)(\xi)\;d\xi,  
\]
with symmetric, positive symbol $\sigma:\R^d\times\R^d\rightarrow \R$.  In \cite{Benassi1997}, the authors define a family of elliptic Gaussian fields $X_{\mathcal A}$ over $\R^d$ through the covariance function given by the integral kernel of the inverse operator $\mathcal A^{-1}$, over the appropriate function space, when it exists. The properties of the field $X_A$ are directly related to those of the pair $(\mathcal A,\mathcal H_{\mathcal A})$. In particular, $X_{\mathcal A}$ has stationary increments if and only if $\sigma$ does not depend on $x$.

\vspace{1em}

Since the set of self-similar Gaussian fields over $\R^d$ with stationary increments is restricted \cite{Dobrushin1979}, the authors in \cite{Benassi1997} instead investigate a local form of self similarity. A random field $X$ is said to be locally asymptotically self-similar of order $H\in (0,1)$ at the point $x_0\in \R^d$ if the limit
\[
\lim_{\rho\rightarrow 0}\frac{1}{\rho^H}(X(x_0+\rho x) - X(x_0))
\]
is non-trivial in law. Evidently, this condition generalizes H-s.s. in Definition \ref{definition:self_similarity_global}. Moreover, by essentially disregarding the low frequency range, it affords the modeler a greater degree of flexibility in adapting the noise field to both the geometry of the underlying index region $D$ as well as enforcing  other more explicit conditions, such as boundary conditions. Standard Brownian motion and the Brownian bridge have the same local scaling properties, for example, although the bridge is not strictly self-similar in the sense of Definition \ref{definition:self_similarity_global}.

\vspace{1em} 

The connection between elliptic Gaussian fields and fractional Brownian surfaces becomes evident when considering the pseudo-differential operator $\mathcal A$ with symbol $\sigma(x,\xi)= \|\xi\|^{d + 2H}$, defined on $\mathcal D_0(\R^d):=\{f\in\mathcal D(\R^d): f(0)=0\}$. If we let $H_A$ be the closure of $\mathcal D_0(\R^d)$ under the inner product $\langle \mathcal Af,g\rangle_{L^2}$, 
then (according to Lemma 1.1., \cite{Benassi1997}) it forms the reproducing kernel Hilbert space for the elliptic Gaussian field  defined by
\begin{equation}\label{eqn:spectral_definition_fbs}
X_{\mathcal A}(x) = \int_{\R^d} \frac{e^{ix\cdot \xi}-1}{\|\xi\|^{\frac{d}{2}+H}}\;d\widehat W(\xi),
\end{equation}
where $\widehat W$ is a complex white noise measure. Apart from a scaling constant, this definition coincides precisely with the spectral characterization of fractional Brownian surfaces \cite{Yaglom1987}. Moreover, by invoking the spectral definition of the fractional Laplacian, we can interpret the action of this integral operator on any function $f$ in the dual Sobolev space $H^{-(\frac{d}{4}+\frac{H}{2})}(\R^d)$, as
\[
\int_{\R^d} \frac{e^{i x\cdot \xi} - 1}{\|\xi\|^{\frac{d}{2}+H}}\hat f(\xi) \;dx = (-\Delta)^{-(\frac{d}{4}+\frac{H}{2})}f(x) - (-\Delta)^{-(\frac{d}{4}+\frac{H}{2})}f(0).
\]
The addition of the term $(-\Delta)^{-(\frac{d}{4}+\frac{H}{2})}f(0)$ ensures that the random field $\{X_{\mathcal A}(x)\}_{x\in\R^d}$ is well-defined. Sample paths of fBs can therefore be formed by solving the fractional Laplace equation with a white noise forcing term, i.e. $X_H(x) = (-\Delta)^{-(\frac{d}{4} + \frac{H}{2})} W(x) - (-\Delta)^{-(\frac{d}{4} + \frac{H}{2})} W(0)$, where $W(x)$ is a white noise field. The covariance of $B_H$ is then given by the integral
\[
C_{X_H}(x,y)=\int_{\R^d} \frac{e^{i (x-y)\cdot \xi}-e^{ix\cdot \xi}-e^{iy\cdot \xi}+1}{\|\xi\|^{\frac{d}{2}+H}}\;d\xi.
\]

\vspace{1em} 

Apart from classical fractional Brownian surfaces, this framework can also be used to define fields corresponding to values of $H\leq 0$, the case $H = 0$ coinciding with the well-known pink noise. The resulting field's sample paths are no longer continuous, but can nevertheless be analyzed in the sense of distributions. Multi-fractional Brownian surfaces represent another interesting class of elliptic fields, whose Hurst parameter is a spatially varying function.

\section{Riesz Fields over Bounded Regions}\label{section:riesz_fields}

%
%
%
%
%
%
The spectral characterization of fractional Brownian surfaces described above reveals how the fractional Laplace operator forms a natural point of departure for defining power law noises over more general index sets. Not only does the operator conform to the geometry of the underlying domain, but it also allows for the imposition of boundary conditions. In \cite{Gelbaum2012}, the author defines the so-called `Riesz fields' over Riemannian manifolds through the use of the Laplace-Beltrami operator. He shows that Riesz fields are generalizations of fractional Brownian surfaces, that they are H\"older continuous with a H\"older coefficient related to the field's Hurst parameter, and that these fields satisfy a form of self-similarity, after accounting for the effect of the Riemannian metric. 

\vspace{1em} 

The fractional Laplace operator has been widely studied in fields such as physics, finance, and hydrology, where it is associated with models of anomalous diffusion. In this paper, we consider homogeneous Dirichlet, Neumann or Robin boundary conditions and treat fractional powers of the Laplace operator in terms of functional calculus related to its spectral decomposition. Indeed, if $D\subset \R^d$ is an open connected domain with piecewise smooth boundary $\partial D$, then the Laplace operator $-\Delta$, subject to the aforementioned boundary conditions, has a discrete, non-negative spectrum with eigenvalues $\{\lambda_k\}_{k=0}^\infty$ satisfying $0\leq \lambda_0 < \lambda_1 \leq \lambda_2 \leq ... \uparrow \infty$ and eigenfunctions $\{\psi_k\}_{k=0}^\infty$ that form a complete basis in $L^2(D)$. The fractional power $(-\Delta)^s$ of the Laplacian $(-\Delta)$, applied to a function $f\in L^2(D)$ can then be expressed as
\[
(-\Delta)^s f(x):= \sum_{k=0}^\infty \lambda_k^s \langle f,\psi_k\rangle \psi_k(x) 
\]
To define Riesz fields over $D$ in accordance with the spectral definition of fractional Brownian surfaces over $\R^d$ requires the fractional inverse $(-\Delta)^{-(\frac{d}{2}+H)}$. For Dirichlet or Robin boundary conditions, the first eigenvalue $\lambda_0$ is strictly positive so that we can define the action of this fractional inverse in terms of the series
\begin{equation}\label{eqn:frac_inverse_kernel}
(-\Delta)^{-(\frac{d}{2} + H)}f(x) = \sum_{k=0}^\infty \lambda_k^{-(\frac{d}{2}+H)}\langle f,\psi_k\rangle \psi_k(x) = \int_D \sum_{k=0}^\infty \lambda_k^{-(\frac{d}{2}+H)}\psi_k(x) \psi_k(y) f(y) \;dy,
\end{equation}
where Fubini's Theorem, together with the uniformly continuous convergence of the series, allows for the interchange of integration and summation. Just as for elliptic Gaussian fields, we now define the covariance function $C_{X_H}:D\times D \rightarrow \R$ of our Riesz field $X_H$ over $D$ as the kernel in \eqref{eqn:frac_inverse_kernel}, i.e.
\begin{equation}\label{eqn:riesz_field_covariance}
C_{X_H}(x,y) = \E{X_H(x)X_H(y)} = \sum_{k=0}^\infty \lambda_k^{-(\frac{d}{2}+H)}\psi_k(x)\psi_k(y)
\end{equation}
Weyl's law, prescribing the asymptotic growth rate of the eigenvalues $\{\lambda_k\}_{k=0}^\infty$ along the order of $O(k^{\frac{2}{d}})$, guarantees that this series converges for any $H\in (0,1)$ and hence $X_H$ is well-defined as a Gaussian field. The Riesz field $X_H$ itself can then be written in the form
\begin{equation}\label{eqn:definition_riesz_field_on_D_dirichlet}
X_H(x,\omega) = \sum_{k=0}^\infty \lambda_k^{-(\frac{d}{4}+\frac{H}{2})} \psi_k(x)Z_k(\omega), \ \text{ where } Z_k \sim N(0,1) \ \text{i.i.d.}.
\end{equation}

\vspace{1em} 

For homogeneous Neumann boundary conditions, $\lambda_0 = 0$ and consequently the fractional inverse $(-\Delta)^{-(\frac{d}{4}+\frac{H}{2})}$ is not defined. Since the corresponding eigenfunction $\psi_0$ is constant, however, the field $X_H$ can be nevertheless be constructed by letting
\begin{equation}\label{eqn:definition_riesz_field_on_D_neumann}
X_H(x) = \sum_{k=0}^\infty \lambda_k^{-(\frac{d}{4}+\frac{H}{2})}(\psi_k(x) - \psi_k(x_0))Z_k,  
\end{equation}
where $x_0\in D$ is some point serving as the origin. The covariance function is defined accordingly. This modification effectively eliminates the zeroth term in the series and therewith the singularity and amounts to imposing $X_H(x_0) = 0$. The same modification appears in the spectral definition \eqref{eqn:spectral_definition_fbs} of fractional Brownian surfaces. Another possibility is to simply leave out the $0^{\mathrm{th}}$ mode, resulting in a field that differs from the one above by an additive constant. In the special case when $D=[0,1]\subset \R$ and $H=\frac{1}{2}$, Definition \eqref{eqn:definition_riesz_field_on_D_dirichlet} amounts to the well-known Fourier expansion of Brownian motion, if Dirichlet boundary conditions are imposed at $t=0$ and Neumann conditions at $t=1$, whereas letting Dirichlet conditions hold at both endpoints represents the Fourier expansion of the Brownian bridge (cf. \cite{Gelbaum2013}).

\vspace{1em}

The Laplace operator's eigenvalues and eigenfunctions depend on the geometry of the region as well as on the imposed boundary conditions, but are invariant under rotations and shifts \cite{Grebenkov2013}. Moreover, when the domain is scaled by a factor $c > 0$ the eigenvalues are rescaled by $1/c^2$, with associated eigenfunctions $\psi_k(x/c)$ for $x\in c D$. In the case of Dirichlet- or Robin boundary conditions, we can relate the covariance function $C_{X_H}^{D}$ over $D$, i.e. the unique integral kernel for $(-\Delta)^{-(\frac{d}{2}+H)}$ over $D$, with the covariance $C_{X_H}^{cD}$ over $cD$, by observing that for any $f\in L^2(c D)$, and $x,y\in c D$,
\begin{align*}
(-\Delta)^{-(\frac{d}{2}+H)}f(x) &= \int_{cD} C_{X_H}^{cD}(x,y) f(y)\;dy =  \sum_{k=0}^\infty \left(\frac{\lambda_k}{c^2}\right)^{-(\frac{d}{2} + H)} \int_{cD}  \psi_k(y/c) f(y)\;dy\; \psi_k(x/c)\\
& = c^{d + 2H}\sum_{k=0}^\infty \lambda_k^{-(\frac{d}{2} + H)}  \int_{D}  \psi_k(\tilde y) f(c \tilde y)c^{-d}\;d\tilde y\; \psi_k(x/c)\\
&= \int_{D} c^{2H} \sum_{k=0}^\infty \lambda_k^{-(\frac{d}{2} + H)}\psi_k(x/c) \psi_k(\tilde y) f(c \tilde y)\;d\tilde y = \int_D c^{2H} C_{X_H}^D(\tilde x,\tilde y) f(c\tilde y)\;d\tilde y,
\end{align*}
where $\tilde x = x/c$ and $\tilde y = y/c$. This implies $C_{X_H}(cx,cy) = c^{2H} C_{X_H}(x,y)$ and therefore that the field $X_H(cx) \ed c^H X_H(x)$ for $x\in D$. Similar scale invariance holds for Riesz fields with Neumann conditions. 

\vspace{1em}

\begin{figure}[ht]
	\centering
	\begin{subfigure}[t]{0.3\textwidth}
	\includegraphics[width=\textwidth]{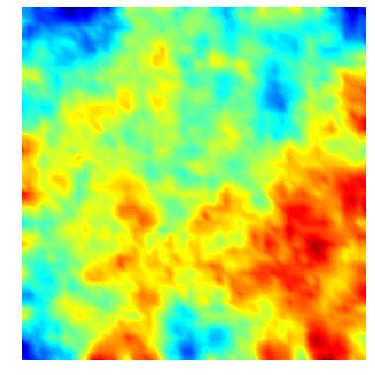} 
	\end{subfigure}%
	\begin{subfigure}[t]{0.3\textwidth}
	\includegraphics[width=\textwidth]{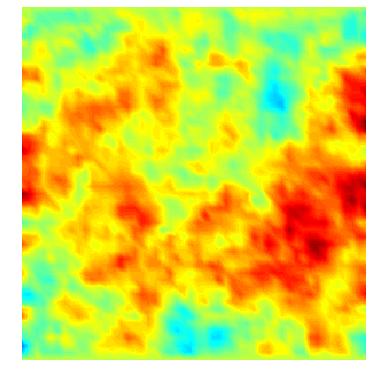}
	\end{subfigure}%
	\begin{subfigure}[t]{0.3\textwidth}
	\includegraphics[width=\textwidth]{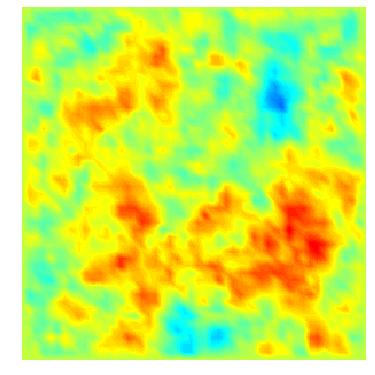}
	\end{subfigure}
	\caption{Realizations of the Riesz field $X_{0.25}$ over a square domain, using the same random seed, but with different boundary conditions.}
	\label{fig:bnd_cond_realizations}
\end{figure}

\begin{figure}[ht]
\centering
	\begin{subfigure}[t]{0.3\textwidth}
	\includegraphics[width=\textwidth]{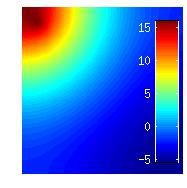}
	\end{subfigure}%
	\begin{subfigure}[t]{0.3\textwidth}
	\includegraphics[width=\textwidth]{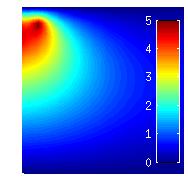}
	\end{subfigure}%
	\begin{subfigure}[t]{0.3\textwidth}
	\includegraphics[width=\textwidth]{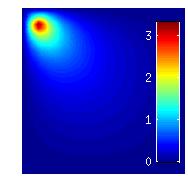}
	\end{subfigure}
	\caption{Covariance functions (scaled by $10^5$) of the Riesz field $X_{0.25}$ at the point $(0.1,0.9)$, corresponding to different boundary conditions}
	\label{fig:bnd_cond_covariance}
\end{figure}

\begin{figure}[H]
\centering
	\begin{subfigure}[t]{0.3\textwidth}
	\includegraphics[width=\textwidth]{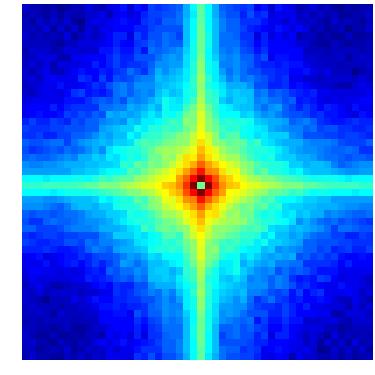}
	\end{subfigure}%
	\begin{subfigure}[t]{0.3\textwidth}
	\includegraphics[width=\textwidth]{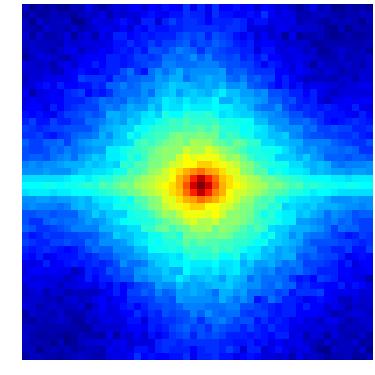}
	\end{subfigure}%
	\begin{subfigure}[t]{0.3\textwidth}
	\includegraphics[width=\textwidth]{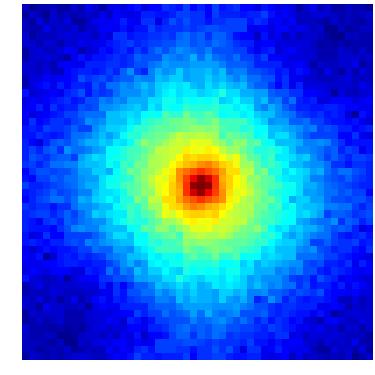}
	\end{subfigure}%
	\caption{Estimated log power spectral densities of the Riesz field $X_{0.25}$ for different boundary conditions.}
	\label{fig:bnd_cond_psd}
\end{figure}

Figures \ref{fig:bnd_cond_realizations} - \ref{fig:bnd_cond_psd} illustrate the effect of boundary conditions on  realizations of the field, its covariance function and its power spectral density. In each case, the leftmost figure corresponds to homogeneous Neumann boundary conditions over the entire boundary, the middle figure corresponds to Neumann conditions on the left and right and Dirichlet conditions at the top and bottom, while the rightmost figure corresponds to homogeneous Dirichlet conditions. Evidently, the field's variance is considerably lower where Dirichlet boundary conditions are enforced, which manifests in both its sample path and covariance (see Figures \ref{fig:bnd_cond_realizations} and \ref{fig:bnd_cond_covariance}). 

\vspace{1em}
  
\begin{figure}[ht]
	\centering
	\begin{subfigure}[t]{0.3\textwidth}
	\includegraphics[width=\textwidth]{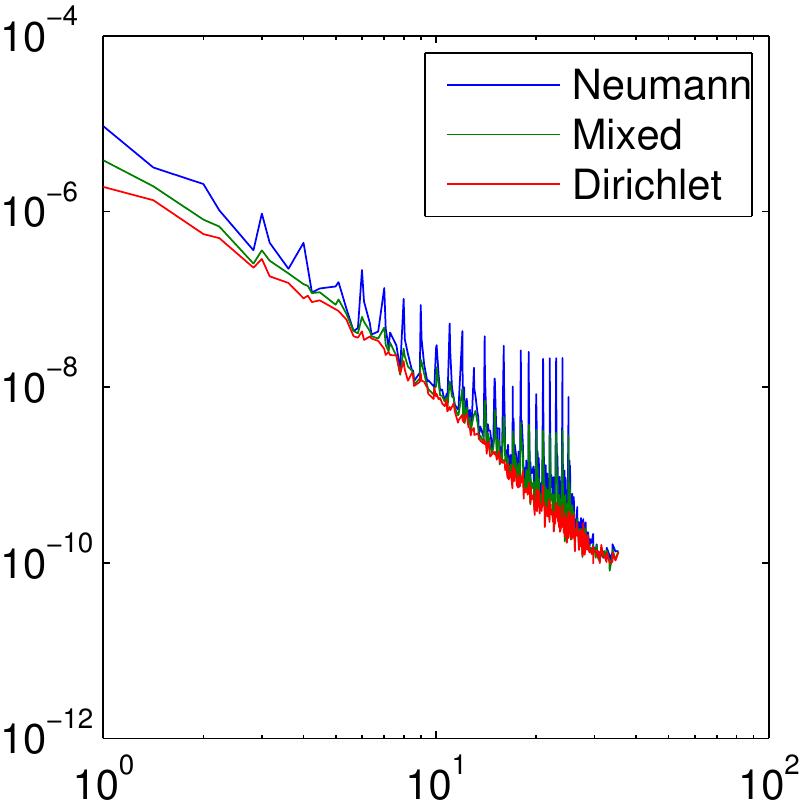}
	\end{subfigure}%
	\hspace{1em}
	\begin{subfigure}[t]{0.3\textwidth}
	\includegraphics[width=\textwidth]{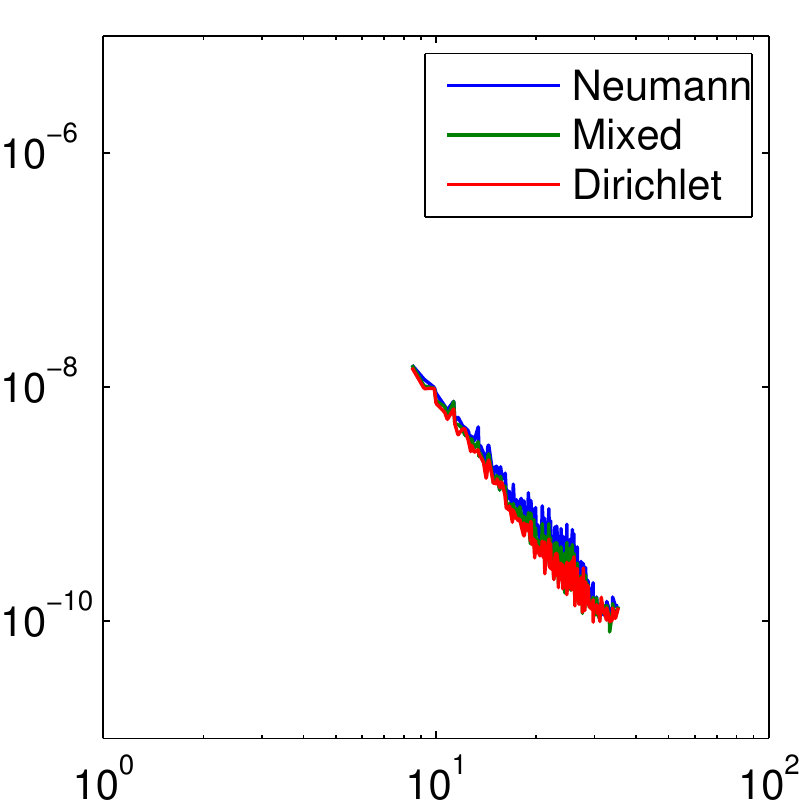}
	\end{subfigure}%
	\caption{Log-log plot of the radial frequency against the azimuthal average of the log power spectral density for different boundary conditions, for all observed frequencies (left) and when frequencies lower than 5Hz are discarded (right).}
	\label{fig:bnd_cond_psd_radial}
\end{figure}

The sample paths, however, seem to have the same degree of `roughness'. This is confirmed by log plots of the appropriate periodograms (see Figures \ref{fig:bnd_cond_psd} and \ref{fig:bnd_cond_psd_radial}), displaying comparable power spectral decay rates in the high frequency regions. 

\section{Numerical Simulations of Riesz Fields} \label{section:samplepaths}


This section discusses three classes of methods for generating numerical simulations of power-law noises on non-standard grids over arbitrary spatial domains. Subsection \ref{subsection:samplepaths_spectral} treats approximations of Riesz field sample paths based on the series representations \eqref{eqn:definition_riesz_field_on_D_dirichlet} and \eqref{eqn:definition_riesz_field_on_D_neumann}, which may require computing the full eigen-decomposition of the discretized Laplace matrix. Subsection    \ref{subsection:samplepaths_fractional} constructs these sample paths as the solution of a fractional-in-space diffusion equation with white noise forcing term. The contour integral method \cite{Hale2008} is a parallelizable algorithm that allows accurate approximations of these solutions to be computed more cheaply than using the eigen-decomposition of the discrete Laplacian. Finally, Subsection \ref{subsection:samplepaths_riesz} treats the generation of sample paths as the integral of the Riesz potential with respect to the white noise measure.

\subsection{Spectral Representation}\label{subsection:samplepaths_spectral}

Equations \eqref{eqn:definition_riesz_field_on_D_dirichlet} and \eqref{eqn:definition_riesz_field_on_D_neumann} express the sample paths of Riesz fields over a bounded domain $D\subset \R^d$ in terms of the eigenvalues and eigenfunctions of the negative Laplacian. These are uniquely determined by the geometry of the underlying domain, together with the imposed boundary conditions. Explicit formulae for them exist over a variety of simple domains, including intervals, hyper-rectangles, parallelepipeds, disks, sectors, spheres and spherical shells, ellipses and elliptical annuli, as well as triangles (see \cite{Grebenkov2013}). In these cases, approximate sample paths $\hat X_H$ can be generated at an arbitrary set of points $\{x_i\}_{i=1}^n$, by truncating the sum in \eqref{eqn:definition_riesz_field_on_D_dirichlet} or \eqref{eqn:definition_riesz_field_on_D_neumann} up to the $K^{\mathrm{th}}$ term, yielding
\[
X_{H,K}^{\eig}(x_i) = \sum_{k=0}^K \lambda_k^{-(\frac{d}{4}+\frac{H}{2})} \psi_k(x_i)Z_k, \ \ \text{for } i = 1,2,...,n.
\]
This spectral approximation has the benefit of being applicable on an arbitrary set of points, in contrast to simulations generated by the circulant embedding technique, where points are required to form a uniformly spaced rectangular grid. Because the Laplacian eigenfunctions are orthogonal in $L^2(D)$, Weyl's law provides an asymptotic error estimate for the spectral approximation of $C_{X_H}$ in terms of the sum of negative powers of the neglected eigenvalues. 
Since eigenfunctions corresponding to high eigenvalues are often highly oscillatory, however, the accuracy of this spectral approximation is conditional on using meshes with a fine enough resolution, so as to avoid additional errors due to aliasing.

\subsubsection{Finite Element Approximations of the Laplace Operator}

In general, the fractional Laplacian must be approximated. This can be done by numerically solving the eigenvalue problem for a matrix representation $A$ of the Laplacian $(-\Delta)$, based on the mesh induced by the nodes and on the given boundary conditions. Various methods are available for discretizing the Laplace operator on a mesh, including finite differences or finite volumes, but we focus here on finite element approximations, since they are well-known for their efficacy in the presence of complex geometries. Let $\{x_i\}_{i=1}^n$ be finite element nodes associated with a regular triangulation $\tau_h = \{\triangle_i\}_{i=1}^m$ of the domain $D$ with maximum mesh spacing parameter $h$, and let $\{\phi_i\}_{i=1}^n$ be the corresponding piecewise polynomial basis functions. The finite element Laplace matrix $A$ can now be formed by letting $A = M^{-1}L$, where $M$ is the mass matrix and $L$ the stiffness matrix with appropriate boundary conditions, defined respectively by
\[
M_{ij} = \int_D \phi_i(x)\phi_j(x)\;dx  \hspace{1em} \text{and} \hspace{1em} L_{ij} = \int_D \nabla\phi_i(x)\cdot\nabla \phi_j(x)\;dx, \hspace{1em} \text{for} \hspace{1em} i,j = 1,2,\hdots n.
\]
The matrix $M$ is strictly positive definite, while $L$ may only be positive semi-definite if pure Neumann boundary conditions are imposed. Both matrices are therefore unitarily diagonalizable with a non-negative spectrum. $A$ on the other hand, is not symmetric in general. However, since $A$ is similar to $M^{1/2}A M^{-1/2} = M^{-1/2} L M^{-1/2}$, which is symmetric semi-positive definite, it is diagonalizable with non-negative spectrum $\sigma(A)$. Therefore, let $A = V\Lambda V^{-1}$, where $\Lambda = \mathrm{diag}(\lambda_1^h, \hdots, \lambda_n^h)$ so that $0\leq \lambda_1^h \leq \lambda^h_2 \leq \hdots \leq \lambda_n^h$, and the columns of $V$ form the coefficients in the finite element approximation of the associated eigenfunctions. If $L$ is strictly positive definite then the sample path $X_H$ can be approximated by means of the spectral expansion 
\[
\widehat X_{H,h}^{\eig}(x) = \sum_{k=1}^n \lambda^{-(\frac{d}{4}+\frac{H}{2})} \psi_k^h(x) Z_k,
\]
where $\psi_k^h(x)= \sum_{i=1}^n V_{ik}\phi_i(x)$ and $\mathbf{Z} = [Z_1,...,Z_n]^T$ is a standard normal random vector in $\R^n$.  


\subsection{Fractional Powers of the Discrete Laplacian}\label{subsection:samplepaths_fractional}

The discussion in Section \ref{section:riesz_fields} suggests that in light of the finite element discretization $A$ of the Laplacian, approximations $\hat X_{H,h}$ of Riesz sample paths can also be obtained by computing the finite element coefficient vector $\hat{\mathbf X}_{H,h}$ as the solution of the discretized fractional diffusion equation, i.e. $A^{\frac{d}{4}+\frac{H}{2}} \hat{\mathbf{X}}_{H,h} = \mathbf{Z}$, where $\mathbf{Z}=[Z_1,...,Z_n]^T$ is a standard normal vector and $A^{-(\frac{d}{4}+\frac{H}{2})} = V \Lambda^{-(\frac{d}{4}+\frac{H}{2})} V^{-1}$. Indeed, the right hand side of the finite element discretization of the Poisson problem with white noise forcing term is given by $M \mathbf{Z}$, since $\int \phi_i\;dW = W(\phi_i)$ for all test functions $\{\phi_i\}_{i=1}^n$, while the stiffness matrix $L$ remains unchanged. The contour integral method presents a more efficient way of computing fractional powers of $A$ without resorting to its full eigen-decomposition.

\subsubsection{The Contour Integral Method}

The contour integral method (CIM) is based on the representation of matrices of the form $f(A)$ as contour integrals around $A$'s spectrum $\sigma(A)$ and allows for the computation of fractional powers of the discrete Laplace operator $A$ without first computing its entire spectrum. Specifically, let $f$ be an analytic function in a region containing $\sigma(A)$ and let $\Gamma$ be a contour lying within this region and winding once around $\sigma(A)$ in a counter-clockwise direction. Then (see \cite{Higham2008}, Definition 1.11 and Theorem 1.12)  
\begin{equation}\label{eqn:fA_contour_integral}
f(A) = \frac{1}{2\pi i} \int_\Gamma f(z) (zI - A)^{-1} \;dz.
\end{equation}
In practice, this matrix-valued integral must be approximated by numerical quadrature, giving rise to the weighted sum
\[
f_N(A) = \sum_{i=1}^N w_i f(\xi_i) (\xi_i I - A)^{-1},
\]
where $w_i$ and $\xi_i$ are a set of weights and nodes. In general, the evaluation of this sum requires computing $n$ resolvent matrices, although these computations can be done completely in parallel. For the purposes of simulating Riesz sample paths, we require only matrix-vector products of the form $A^{-(\frac{d}{4}+\frac{H}{2})}\mathbf{Z}$, where $\mathbf{Z}$ is the standard normal vector, provided $A$ is non-singular. Since $A = M^{-1}L$, the sum above takes the form
\[
\hat{\mathbf{X}}_{H,h}^{\cim} := A^{-(\frac{d}{4}+\frac{H}{2})}\mathbf{Z} \approx \sum_{i=1}^N w_i \xi_i^{-(\frac{d}{4}+\frac{H}{2})} (\xi_i M - L)^{-1}M \mathbf{Z},
\]
requiring $N$ system solves, so that neither $A$ nor its inverse need be formed explicitly. To compute the associated covariance matrix, however, the full matrix inverse is required. In the case of pure Neumann boundary conditions, it was shown (c.f. Theorem 4.1 \cite{Burrage2012}) that $A_{\mathrm{neu}}^{-( \frac{d}{4} + \frac{H}{2})}:= V_{\mathrm{neu}} \Lambda^{-( \frac{d}{4} + \frac{H}{2})} V_{\mathrm{neu}}^{-1}$, where $\Lambda_{\mathrm{neu}} = \mathrm{diag}(\lambda_2^h,...,\lambda_n^h)$ and $V_{\mathrm{neu}}$, is formed from all but the first columns of $V$, can be computed by the contour integral 
\[
A_{\mathrm{neu}}^{-( \frac{d}{4} + \frac{H}{2})} = \frac{1}{2\pi i} \int_{\Gamma_2} z^{-( \frac{d}{4} + \frac{H}{2})} (zI - A)^{-1} \;dz,
\]
where $\Gamma_2$ contains $\lambda_2,...,\lambda_n$, but not $\lambda_1$.
\vspace{1em}

Unfortunately, the accuracy of conventional approaches, such as applying the trapezoidal rule to the circular contour enclosing $\sigma(A)$, deteriorates as the condition number $\kappa(A)$ of $A$ grows, with a convergence rate that depends linearly on $\kappa(A)$. For the finite element Laplacian, $\kappa(A) = \lambda_n^h/\lambda_1^h$ in turn grows as the mesh is refined. In \cite{Hale2008}, the authors develop a numerical quadrature scheme whose accuracy deteriorates only logarithmically in terms of $\kappa(A)$ for functions $f$ that are analytic in the slit complex plane $\mathbb C\backslash  (-\infty,0]$, by first mapping the region $\mathbb C\backslash ( (-\infty,0] \cup \sigma(A))$ conformally onto an annulus and applying the trapezoidal rule there. It can then be shown (see \cite{Hale2008} Theorem 2.1) that $\|f(A) - f_N(A)\| = O(e^{-\pi^2 N/(\log(\kappa(A) + 3)})$. This quadrature scheme was used in \cite{Burrage2012} to compute the fractional FEM Laplace operator $A$ in aid of approximating the solution of a fractional-in-space reaction diffusion equation. 

\vspace{1em}

\begin{figure}[H]
\centering
\begin{subfigure}[t]{0.25\textwidth}
\includegraphics[width=\textwidth]{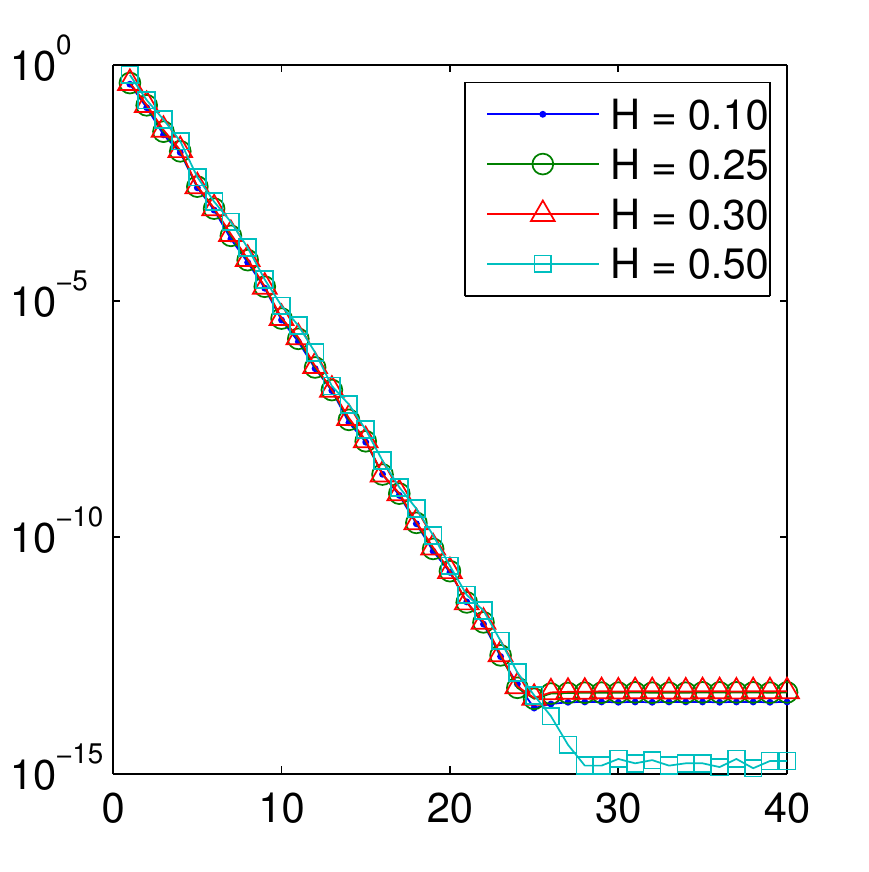}
\end{subfigure}%
\begin{subfigure}[t]{0.25\textwidth}
\includegraphics[width=\textwidth]{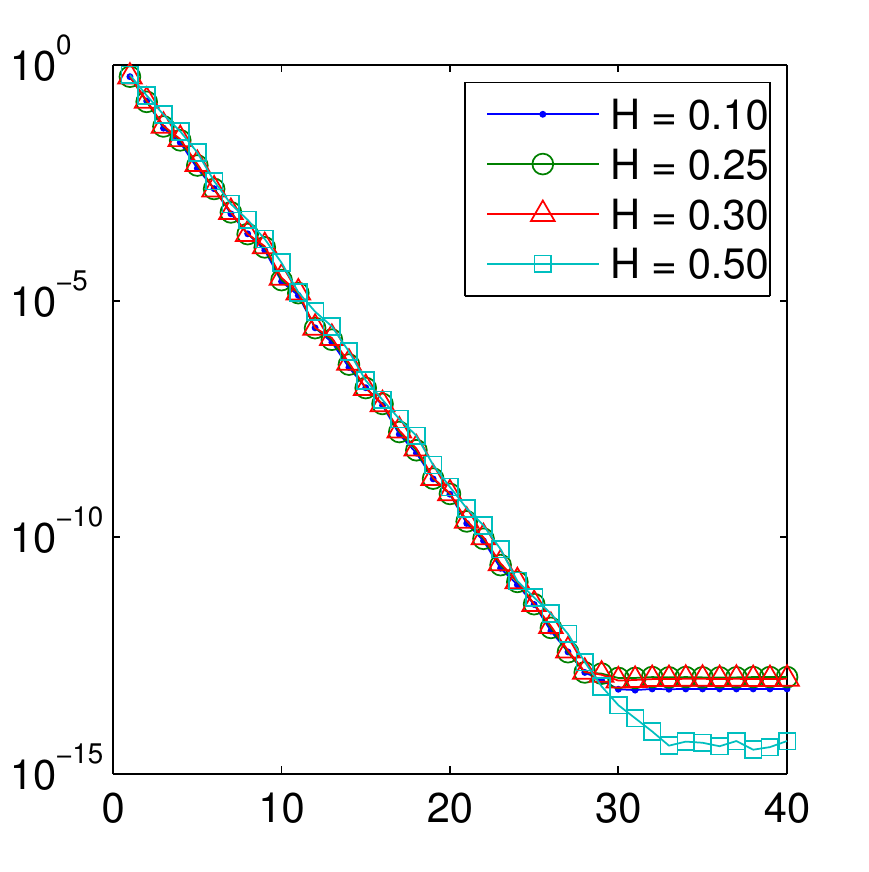}
\end{subfigure}%
\begin{subfigure}[t]{0.25\textwidth}
\includegraphics[width=\textwidth]{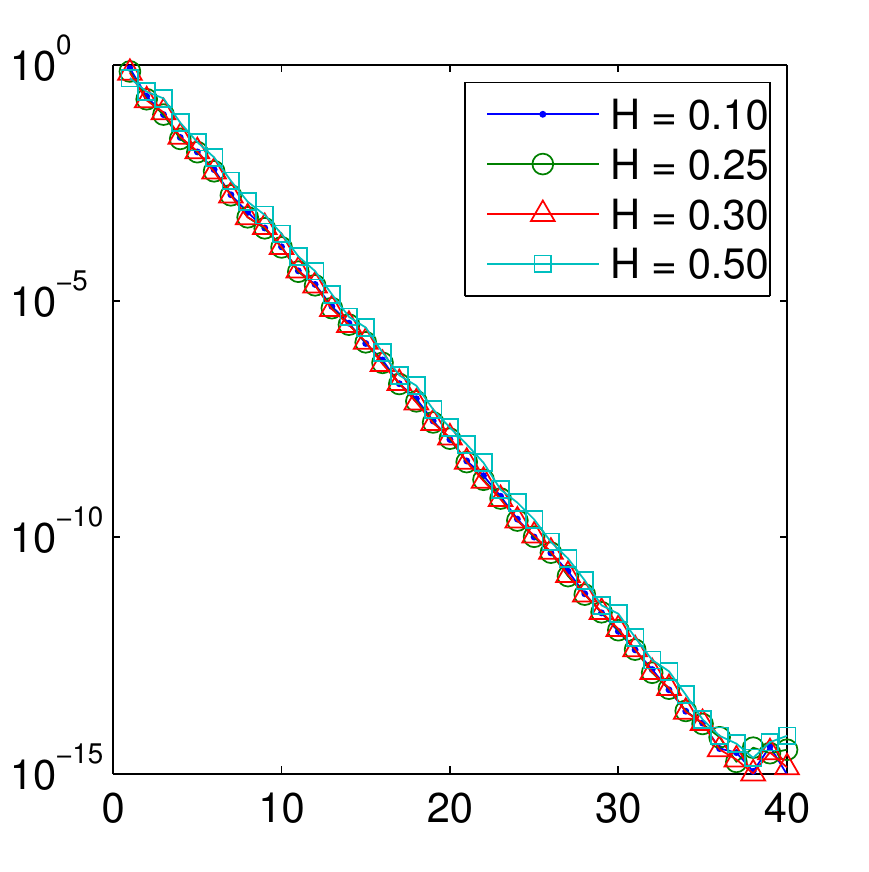}
\end{subfigure}%
\begin{subfigure}[t]{0.25\textwidth}
\includegraphics[width=\textwidth]{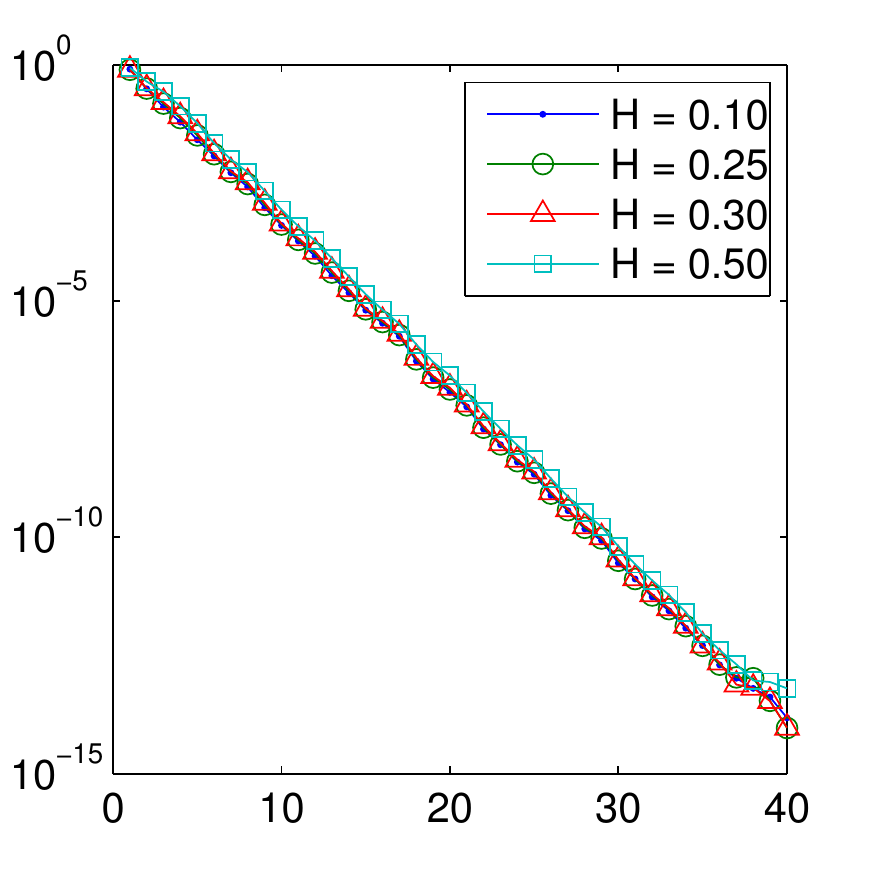}
\end{subfigure}
\caption{The relative $L^\infty$-error of the CIM approximation of the sample path $\mathbf{X} = A^{-\frac{d}{4}+\frac{H}{2}}\mathbf{Z}$ for successive refinements of the spatial mesh.}
\label{fig:cim_convergence}
\end{figure}

To investigate the accuracy and efficiency of the CIM method, we compute the finite element coefficients $\hat{\mathbf{X}}_{H,h}^{\cim}$ of a single sample path of the Riesz field satisfying homogeneous Dirichlet conditions for different values of the Hurst parameter, different levels of mesh refinement, and different numbers of quadrature nodes. The quadrature nodes and weights were computed using Algorithm 1 in \cite{Burrage2012} (see also \texttt{method1.m} in \cite{Hale2008}), based on Driscoll's Schwarz-Christoffel Toolbox \cite{Driscoll1996}. For the first and second mesh refinement levels, the reference path was computed using Matlab's \texttt{mpower} command and the backslash operation `$\backslash$'. For the finer meshes, we used the CIM method with 100 quadrature points. Figure \ref{fig:cim_convergence} shows the convergence rate of the CIM method for different meshes. As expected, the method converges exponentially, but the convergence rate deteriorates as the mesh becomes finer, giving rise to a higher condition number. 

To assess the efficiency of the CIM, we compare its CPU time with that of forming the fractional inverse of $A$ by computing its full eigen-decomposition, using the \texttt{eig} function in Matlab. We ran our computations on a Intel Core i5-2520M CPU @ 2.50GHz x 4, running Matlab R2012 without parallelization. Table \ref{table:cim_vs_eig} clearly shows the advantage in computational cost of the CIM over using the eigen-decomposition of $A$. For a comparable (or even better) relative error, the CIM is an order of magnitude faster, especially for finer meshes.

\begin{table}[H]
\centering
	\begin{tabular}{|c||c|c|c|c|c|}
		\hline
		Level & Nodes & $\kappa(A)$ & CIM time & eig time & eig error\\
		\hline
		1 & 144 & 148 & 0.1056 & 0.0068 & 2.2629e-14\\ 
		\hline 
		2 & 529 & 623 & 0.3449 & 0.2196	&  6.3948e-14\\ 
		\hline 
		3 & 2025 & 2525 & 1.4320 & 11.5975 & 4.6633e-13\\ 
		\hline 
		4 & 7921 & 10133 & 10.3513 & 645.7518 & 1.0503e-11\\ 
		\hline 
	\end{tabular} 
	\caption{Condition numbers and computational times for both the CIM and the eigenvalue method for different spatial refinement levels.}
	\label{table:cim_vs_eig}
\end{table}

\subsection{Discretization of the Riesz Kernel}\label{subsection:samplepaths_riesz}

In this section we propose a method for simulating power law noises as convolutions of the Riesz potential \eqref{eqn:riesz_potential} with white noise fields, as an alternative to using the inverse fractional Laplace operator. This formulation does not allow for the explicit enforcement of boundary conditions, although it is simpler to implement than previously discussed methods, since it does not require the inversion of fractional Laplace operators. To prevent fields from exhibiting spurious correlations over non-convex domains (see Figure \ref{figure:covariance_geometry}), we replace the Euclidean distance appearing in the Riesz kernel by the distance $d_D(x,y)$ of the shortest path in $D$ between the points, i.e. 
\begin{equation}
d_D(x,y) := \min\{ \mathrm{length}(\gamma): \gamma \text{ is a path from $x$ to $y$}\},
\end{equation}
giving rise to the modified kernel $k_H(x,y):= d_D(x,y)^{-\frac{d}{2}+H}$. At any point $x\in D$, the value of the random field $X_H^{\riesz}(x)$ is then given by
\[
X_H^{\riesz}(x) = c_{H+\frac{d}{2}}\int_D k_H(x,y) \;dW(y), 
\] 
where $c_{H+\frac{d}{2}}$ is given in \eqref{eqn:riesz_potential} and $W$ is the white noise measure defined in Section \ref{section:preliminaries}. 

\vspace{1em} 

\begin{figure}[ht]
\centering
\begin{subfigure}[t]{0.35\textwidth}
\includegraphics[width=\textwidth]{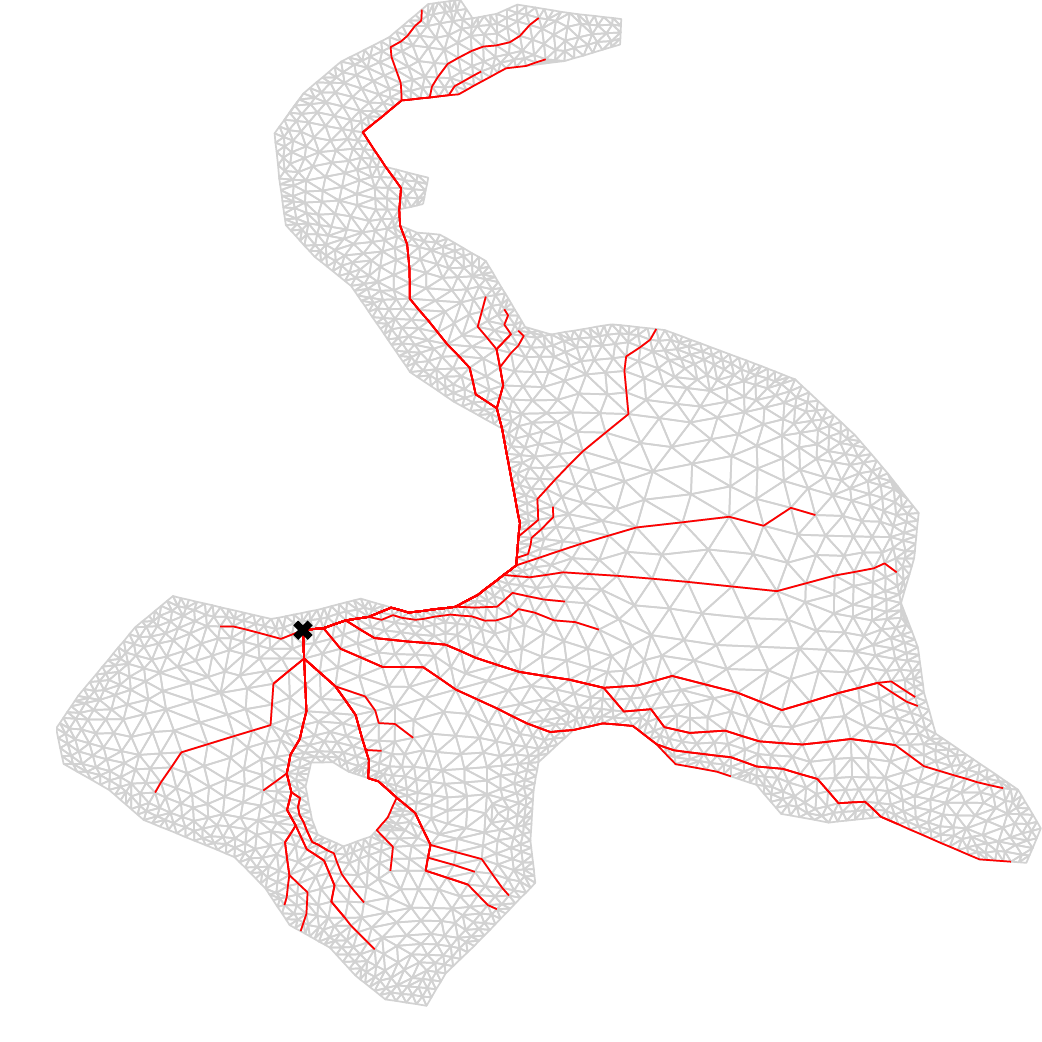}
\caption{Shortest paths from point marked by x to 40 other random points in the domain.}
\end{subfigure}%
~
\begin{subfigure}[t]{0.35\textwidth}
\includegraphics[width=\textwidth]{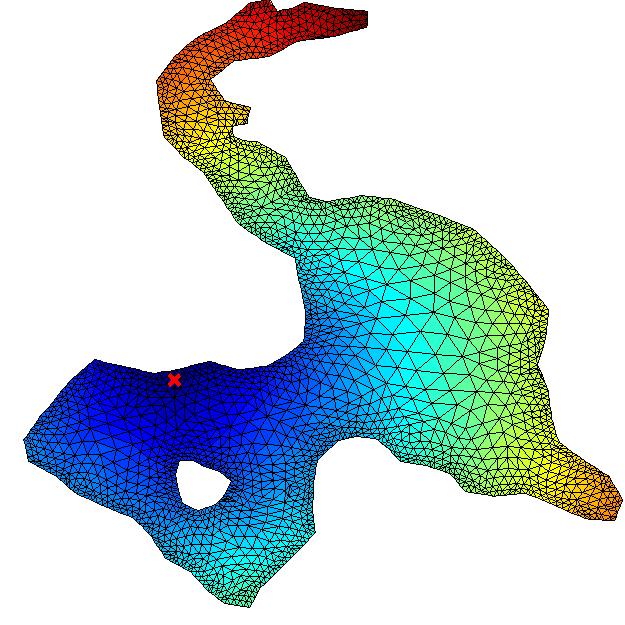}
\caption{Distances from point marked by x to all other points in the domain}
\end{subfigure}%
\caption{Illustrations of the modified metric over the lake domain.}
\label{fig:riesz_potential_distances}
\end{figure}

As before, let $\tau_h = \{\triangle_i\}_{i=1}^m$ be a regular triangulation of the region $D$. For any $x\in D$, we approximate the integral kernel $k_H(x,\cdot)$ by the piecewise constant function 
\[
\hat k_H(x,y) := \sum_{i=1}^m k_H(x,y_i^*)\mathbbm{1}_{\triangle_i}(y),
\] 
where $\mathbbm 1$ is the indicator function and $y_i^*\in \triangle_i$ is a representative point in the interior of the $i^{th}$ element $\vartriangle_i$. In our computations, we take $y_i^*$ to be the centroid of the $i^{th}$ element and compute the shortest distance between any two finite element nodes by means of Floyd's algorithm \cite{Floyd1962} (see Figure \ref{fig:riesz_potential_distances}). 

\vspace{1em}

\begin{figure}[ht]
\centering
\begin{subfigure}[t]{0.35\textwidth}
\includegraphics[width=\textwidth]{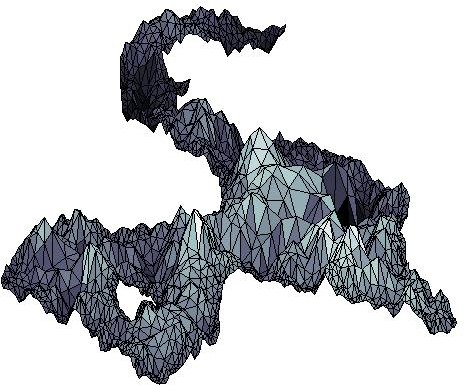}
\caption{Sample path of the power-law noise $\hat X_{0.25,h}^{\riesz}$ over the lake region.}
\label{subfig:riesz_potential_sample}
\end{subfigure}%
~
\begin{subfigure}[t]{0.3\textwidth}
\includegraphics[width=\textwidth]{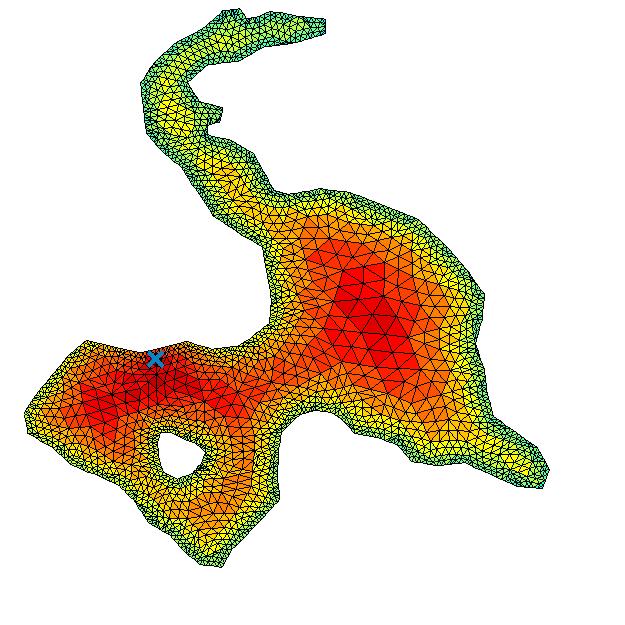}
\caption{Contribution of each noise term to the variance at the point marked by `x'.}
\label{subfig:riesz_potential_variance}
\end{subfigure}%
~
\begin{subfigure}[t]{0.3\textwidth}
\includegraphics[width=\textwidth]{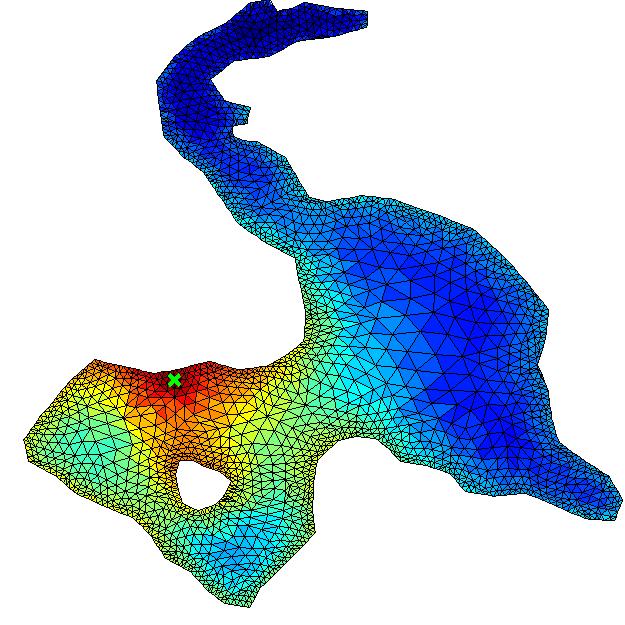}
\caption{The correlation at the point marked by `x'.}
\label{subfig:riesz_potential_correlation}
\end{subfigure}%
\caption{Sample paths and point-wise covariance structure of $\hat X_{0.25,h}^{\riesz}$.}
\label{fig:riesz_potential_field}
\end{figure}

\noindent Consequently, the random field $X_H^{\riesz}$ can be approximated by 
\begin{equation}\label{eqn:riesz_potential_field} 
\hat X_{H,h}^{\riesz}(x) := \sum_{i=1}^m k_H(x,y_i^*) W(\triangle_i) = \sum_{i=1}^m k_H(x,y_i^*) |\triangle_i|Z_i, \ \text{with } Z_i \sim N(0,1) \ \mathrm{i.i.d.}.
\end{equation}
Like Hosking's model, this formulation expresses the field at a given point $x\in D$ as a linear combination of white noise sources located in each of the elements, so that both the size of the element and the distance from $x$ to the element's centroid determine the strength of the influence of the noise source. For points lying close together, the field is largely determined by the same random disturbances leading to a higher level of correlation, the strength of which depends on the decay rate of the kernel, i.e. on $H$. Figure \ref{subfig:riesz_potential_correlation} shows the correlation of the field $\hat X_{0.25,h}^{\riesz}$, whose sample path is depicted in Figure \ref{subfig:riesz_potential_sample}, at a given point. As expected, the correlation decreases as we move away from the point. Unlike elliptic Gaussian fields with Neumann boundary conditions (see Figure\ref{fig:bnd_cond_covariance}), however, the correlation is always positive. An unusual property of this field is that the correlation of the point `x' with points on the boundary is slightly higher than with points that are closer but that lie in the interior. This is due to the fact that points on the boundary are influenced by fewer noise sources than their interior neighbors, rendering them more correlated. Figure \ref{subfig:riesz_potential_variance} shows that both the distance of the noise source from a point $x$, as well as the size of the element determine its contribution to the field's variance at $x$.

\subsubsection*{Spatially Varying Hurst Parameters}

\begin{figure}[ht]
\centering
\begin{subfigure}[t]{0.45\textwidth}
\includegraphics[width=\textwidth]{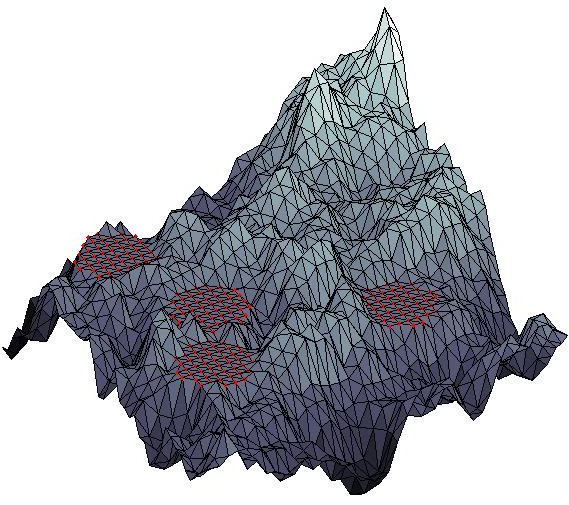}
\caption{$\hat X_{H,h}^{\cim}$ with homogeneous Dirichlet conditions on the red circles.}
\label{subfig:fbm2d_plus_interior_bnd}
\end{subfigure}%
~
\begin{subfigure}[t]{0.45\textwidth}
\includegraphics[width=\textwidth]{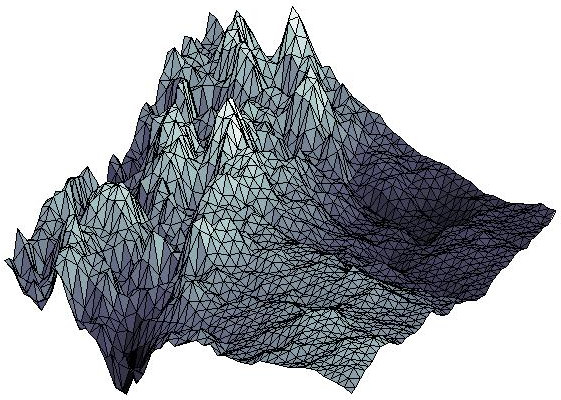}
\caption{$\hat X_{H,h}^{\riesz}$ with Hurst parameter\\ $H(x,y) = 0.5 + 0.5/\pi*\arctan(10*(x-0.5))$.} 
\label{subfig:fbm2d_plus_Hx}
\end{subfigure}
\caption{Sample paths of random fields generated using the Riesz potential.}
\end{figure}

Although $\hat X_{H,h}^{\riesz}$ cannot be used when it is necessary to enforce boundary conditions, the explicit appearance of the Hurst parameter in Equation \eqref{eqn:riesz_potential_field} allows us to model fields whose Hurst parameter $H$ is spatially varying, by letting 
\[
\hat X_{H,h}^{\riesz}(x) := \sum_{i=1}^m d_D(x,y_i^*)^{-\frac{d}{2}+H(x)} W(\triangle_i).
\] 
Figure \ref{subfig:fbm2d_plus_Hx} shows the sample path of a field $\hat X_{H,h}^{\riesz}(x)$ with Hurst parameter that increases along the x-direction from left to right.

\section{Conclusion}\label{section:conclusion}

Despite claims to the contrary, statistical self-similarity seems not to be a universal law of nature, but rather a set of descriptive properties, observable in an variety of guises in physical- biological and man-made systems. Due to the  multitude of ways in which self-similarity can be manifested, especially over general, multi-dimensional index sets, statistical models of self-similar or power-law noises should be sufficiently flexible  to accommodate both constraints imposed by observations, and the needs of the modeler. In this paper we showed that the elliptic Gaussian fields form a wide class of locally self-similar random fields of known smoothness that can be generated over non-standard spatial domains and on arbitrary meshes. We proposed three algorithms for generating numerical simulations of power-law noise and discussed and compared their properties, strengths and limitations.

\vspace{1em}

The theory of Gaussian models for spatially varying power-law noises, also known as elliptic Gaussian processes \cite{Benassi1997} \cite{Benassi2003}, or Riesz fields \cite{Gelbaum2012,Gelbaum2013}, is fairly recent and there is a need for a more complete understanding of the nature of solutions of SDE's and SPDE's and their approximations, when  the underlying parameters that are spatially varying power-law noises. This includes a convergence theory for approximations based on finite elements, or finite differences. Another direction of future research involves the quantification of more specific features of the field, such as the presence of a grain, as well as the incorporation of these into the random field model.

\bibliography{pink_noise_bib}
\bibliographystyle{siam}
\end{document}